\documentclass{pasj00}
\draft
\begin{document}
\SetRunningHead{H. Kawahara et al.}
{Soft X-ray Transmission Spectroscopy of WHIM}
\title{Soft X-ray Transmission Spectroscopy 
of Warm/Hot Intergalactic Medium
with XEUS}
\author{
Hajime \textsc{Kawahara}\altaffilmark{1},
Kohji \textsc{Yoshikawa}\altaffilmark{1},
Shin \textsc{Sasaki}\altaffilmark{2},
Yasushi \textsc{Suto}\altaffilmark{1},\\
Nobuyuki \textsc{Kawai}\altaffilmark{3},
Kazuhisa \textsc{Mitsuda}\altaffilmark{4},
Takaya \textsc{Ohashi}\altaffilmark{2}, and
Noriko Y. \textsc{Yamasaki}\altaffilmark{4}
}

\altaffiltext{1}{Department of Physics, School of Science, 
The University of Tokyo, Tokyo 113-0033}
\email{kawahara@utap.phys.s.u-tokyo.ac.jp}

\altaffiltext{2}{Department of Physics, Tokyo Metropolitan University,
 \\ 1-1 Minami-Osawa, Hachioji, Tokyo 192-0397}

\altaffiltext{3}{Department of Physics, Tokyo Institute of Technology,
 \\ 2-12-1 Ookayama, Meguro-ku, Tokyo 152-8551}

\altaffiltext{4}{The Institute of Space and Astronautical Science
(ISAS), \\ Japan Aerospace Exploration Agency (JAXA), 
\\ 3-1-1 Yoshinodai, Sagamihara, Kanagawa 229-8510}

\Received{2005/04/27}
\Accepted{2005}

\KeyWords{cosmology: miscellaneous --- X-rays: general --- 
methods: numerical}

\maketitle

\begin{abstract}
We discuss the detectability of Warm/Hot Intergalactic medium (WHIM) via
the absorption lines toward bright point sources with a future X-ray
satellite mission, {\it XEUS}.
While we consider bright QSOs as specific examples, the methodology 
can be applied to bright gamma-ray burst afterglows.
We create mock absorption spectra for
bright QSOs (more than 20 QSOs over the all sky)
using a light-cone output of a
cosmological hydrodynamic simulation. We assume that WHIM is under
collisional and photo-ionization equilibrium.  If WHIM has a constant
metallicity of $Z=0.1Z_\odot$, approximately 2 O{\sc vii} absorption
line systems with $>3\sigma$ will be detected on average along a random
line-of-sight toward bright QSOs up to $z=0.3$ for
a 30ksec exposure.  
\end{abstract}

%\clearpage
\section{Introduction}

More than 50 percent of cosmic baryons is dark \citep{Fukugita1998,
Fukugita2004}. The abundance, physical properties, and spatial
distribution of such dark baryons constitute one of the most important
unsolved problems in observational cosmology. Recent numerical
simulations suggest that a major fraction of dark baryons takes a form
of WHIM (Warm/Hot Intergalactic medium) roughly with temperature $10^5
{\rm K}< T < 10^7{\rm K}$ and over-density $ 1< \delta< 1000$
\citep{Cen1999a,Dave2001}.  WHIM traces the large-scale filamentary
structure of mass (dark matter) distribution more faithfully than hot
intracluster gas ($T > 10^7 {\rm K}$ typically) and galaxies both of
which preferentially reside in clusters that form around the knot-like
intersections of the filamentary regions (e.g., Suto et al. 2004a,
2004b; Kang et al. 2005). This implies that WHIM carries important
cosmological information in a complementary fashion to distributions of
galaxies in optical and of clusters in X-ray.

{\it DIOS} (Diffuse Intergalactic Oxygen Surveyor) is a dedicated soft
X-ray mission which aims at unveiling dark baryons in the universe
\citep{Ohashi2004} via oxygen emission from WHIM. Detectability of such
emission features was examined in detail both for mock blank survey
(\cite{Yoshikawa2003}; hereafter Paper I) and for mock targeted survey
of outskirts of clusters in the local universe (\cite{Yoshikawa2004};
hereafter Paper II).  These authors find that approximately 20 percent
of cosmic baryons leaves emission signatures detectable by {\it
DIOS}. This fraction corresponds to a relatively higher temperature tail
($10^6 {\rm K}< T < 10^7{\rm K}$) of the entire WHIM distribution, and
is sensitive to the assumed metallicity of WHIM. They adopt a
metallicity model proposed by \citet{Cen1999b} and \citet{Aguirre2001},
but the metallicity of WHIM is admittedly very uncertain at this moment.

Another complementary strategy of detecting elusive WHIM is to search
for its absorption signatures. In fact, several independent groups
reported successful detections of absorption lines of highly ionized
metals in spectra of background QSOs and clusters; O{\sc vii} and O{\sc
viii} \citep{Fang2002a}, O{\sc vi} at $z \sim 0.2$ in the {\it FUSE}
spectrum of QSO H1821+643 \citep{Tripp2000}, and more recently O {\sc
vii}, N {\sc vii} and N{\sc vi} along the line of sight toward the
Blazar Mkn 421 \citep{Nicastro2005,Nicastro2005b}.
See also \citet{Nicastro2002,Fang2002b,Mathur2003,Fujimoto2004}.  In
particular \citet{Nicastro2005} estimate $\Omega_{\rm b}^{\rm WHIM} =
(2.7 {+3.8 \atop -1.9}\%) \times 10^{-[{\rm O}/{\rm H}]_{-1}}$ adopting
a temperature of $10^{6.1}$K for the WHIM, where $10^{[{\rm O}/{\rm
H}]_{-1}}$ is the oxygen-to-hydrogen number density ratio in units of
$0.1$ times the solar value. While the above value seems to be in good
agreement with that expected from the previous simulations
\citep{Cen1999a,Dave2001}, the estimate is still sensitive to the
assumed temperature and metallicity of the WHIM along the single line of
sight to Mkn 421 even apart from its limited statistics.

This paper examines the detectability of WHIM via absorption, instead of
emission, lines using the same cosmological hydrodynamic simulation data
as Paper I. 
In particular we consider QSOs
as a bright X-ray lighthouse, and create mock absorption
spectra for a future X-ray satellite mission, {\it XEUS} (X-ray
Evolving-Universe Spectroscopy). 
This methodology can be equally applied to gamma-ray burst (GRB) X-ray
afterglows as first proposed by \citet{Fiore2000}.

The rest of the paper is organized as follows; section 2 describes our
methodology and model assumptions in computing mock transmission spectra
of the QSOs through WHIM. 
We present examples of mock spectra
and their basic statistics in \S 3. We discuss the detectability of WHIM
via absorption in \S 4 and also consider how to extract physical
properties of WHIM.
Finally section 5 is devoted to the conclusions of the paper and to
further discussion.

\section{Creating Mock Transmission Spectra of WHIM toward QSOs
from Cosmological Hydrodynamic Simulations}

We compute mock transmission spectra of WHIM using the same simulation
data as Paper I.  We create a stacked light-cone output for $0<z<0.3$
from the simulation data. Then we consider 6400 different line-of-sight
directions, and compute mock spectra for a bright QSO
taking into account the cosmic X-ray background (CXB) and UV background
(UVB) for photo-ionization sources.  Our methodology is described in the 
following subsections.

\subsection{Cosmological Hydrodynamic Simulation}

We use the data of \citet{Yoshikawa2001} who performed a simulation with
a hybrid code of Particle--Particle--Particle--Mesh (PPPM) Poisson
solver and smoothed particle hydrodynamics (SPH).  Both dark matter and
gas employ $128^3$ particles within a periodic simulation cube of
$L_{\mathrm{box}} =75h^{-1}\mbox{Mpc}$ per side (in comoving). They
adopt a standard ${\mathrm{\Lambda CDM}}$ model in which the density
parameter $\Omega_{\rm m} = 0.3$, the baryon density parameter $\Omega_b
= 0.015 h^{-2}$, the dimensionless cosmological constant $\Omega_\Lambda
= 0.7$, the rms density fluctuation smoothed over a scale of $8 h^{-1}
\mathrm{Mpc}$, $\sigma_8=1.0$, and the Hubble constant in units of
$\mathrm{100km/s/Mpc}$ $h = 0.7$. The simulation includes the effect of
radiative cooling, but the energy feedback from supernova and the UV
background heating are neglected.  The simulation that we
used here was performed in 2000. This is why the adopted value of
$\Omega_b=0.03$ is $\sim 30$\% smaller than more recent estimate, for
instance, $\Omega_b = (0.024\pm 0.001) h^{-2}$ \citep{Spergel03}.  In
this sense our results may underestimate the real absorption
signatures by the corresponding amount.

We create a light-cone output for $0<z<0.3$ following the procedure
described in Paper I. The maximum value of $z=0.3$ is adopted so as to  
ensure that the O{\sc viii} line is located in a spectrum region
relatively free from the Galactic confusion (Paper I), while this may be
very conservative.  The output extends a $5^\circ \times 5^\circ$ region
on the sky, which corresponds to the simulation boxsize at $z=0.3$;
$L_{\rm box}/(1+0.3)/d_{\rm A}(z=0.3) \approx 5^\circ$ with $d_{\rm
A}(z)$ being the angular diameter distance at redshift $z$.

\subsection{Calculation of Optical Depths}

Consider one particular line of sight (LOS) with its unit direction
vector being ${\bf \hat n}$. We divide the LOS into 5000 bins $x_i$
($i=1 \sim 5000$) so that their bin width is equal in comoving scale,
i.e., $x_{i+1}-x_i=d_{\rm C}(z=0.3)/5000$, where $d_{\rm C}(z)$ is the
comoving distance at redshift $z$.  The gas density $\rho(x_i) $,
temperature $T(x_i)$ and peculiar velocity $v_{r}(x_i)$ of the
intergalactic medium (IGM) at the $i$-th bin are estimated by gathering
contribution from all the simulation gas particles:
%%-----------------------------------------------------------------------
\begin{eqnarray}
    \rho(x_i) &=& \sum^N_{j=1} m \, W (x_i{\bf \hat n}- {\bf r}_j,h_j), \\
    T(x_i)    &=& 
    \, \sum^N_{j=1} m \, \frac{T_j}{\rho_j} 
\, W (x_i{\bf \hat n}- {\bf r}_j,h_j), \\
    v_{r} (x_i) &=& 
     \sum^N_{j=1} m \, \frac{v_{r,j}}{\rho_j} 
\,  W (x_i{\bf \hat n}- {\bf r}_j,h_j).
\end{eqnarray}
%%-----------------------------------------------------------------------
In the above expressions, $j$ denotes an index for gas particles ($j=1
\sim N$), $T_j$ , $v_{r,j}$, ${\bf r}_j$, $h_j$, and $\rho_j$ are
temperature, peculiar velocity, position vector, smoothing length, and
density of the $j$-th gas particle, and $m$ is the mass of gas particles 
(independent of $j$).  \citet{Yoshikawa2001} classify some fraction of
gas particles as ``galaxies'' if they satisfy both the Jeans condition
and the over-density criterion. Those particles are excluded from the
above summation.  We adopt the smoothing kernel which is identical to
that employed in the simulation \citep{Yoshikawa2001}:
%%-----------------------------------------------------------------------
\begin{eqnarray}
     W({\bf x}-{\bf r},h) = \frac{1}{\pi h^3} 
      \left\{
       \begin{array}{lr}
	1 - (3/2) u^2 + (3/4)u^3 & \mbox{if $0 \le u \le 1$} \\
	(2 - u)^3/4 & \mbox{if $1 \le u \le 2$} \\
	0 & \mbox{otherwise} \\
       \end{array}
      \right.,
\end{eqnarray}
%%-----------------------------------------------------------------------
where $u \equiv |{\bf x}-{\bf r}|/h$. 

We estimate the number density of an ionized metal $A^{n+}$ (with charge
{\it n}+) at the $i$-th bin as
%%-----------------------------------------------------------------------
\begin{equation}
    n_{A^{n+}} (x_i) = \sum^N_{j=1} F_{A^{n+}}(T_j,\rho_j)  
        \left(\frac{n_{A}}{n_H}\right)_j  
                      \frac{X}{m_p} m \,  W (x_i{\bf \hat n}- {\bf
		      r}_j,h_j), 
\end{equation}
%%-----------------------------------------------------------------------
where $X$ is the hydrogen mass fraction (we adopt 0.755), 
$m_p$ is the proton mass, 
and $Z_{A,j}$ and $F_{A^{n+}} (T_j,\rho_j)$
represent the metallicity of $A$ and the ionization fraction of $A^{n+}$
of the $j$-th gas particle.  In most cases, we assume a
constant metallicity for simplicity:
%%-----------------------------------------------------------------------
\begin{equation}
   Z_{A,j}=0.1 Z_{A,\odot},
\end{equation}
%%-----------------------------------------------------------------------
or equivalently a constant number ratio:
%%-----------------------------------------------------------------------
\begin{equation}
\label{eq:na2nh}
   (n_{A}/n_H)_j = 0.1 (n_{A}/n_H)_\odot
\end{equation}
%%-----------------------------------------------------------------------
independently of the time and the local density (see \S 3.1 below for
other models).

\begin{figure}[htbp]
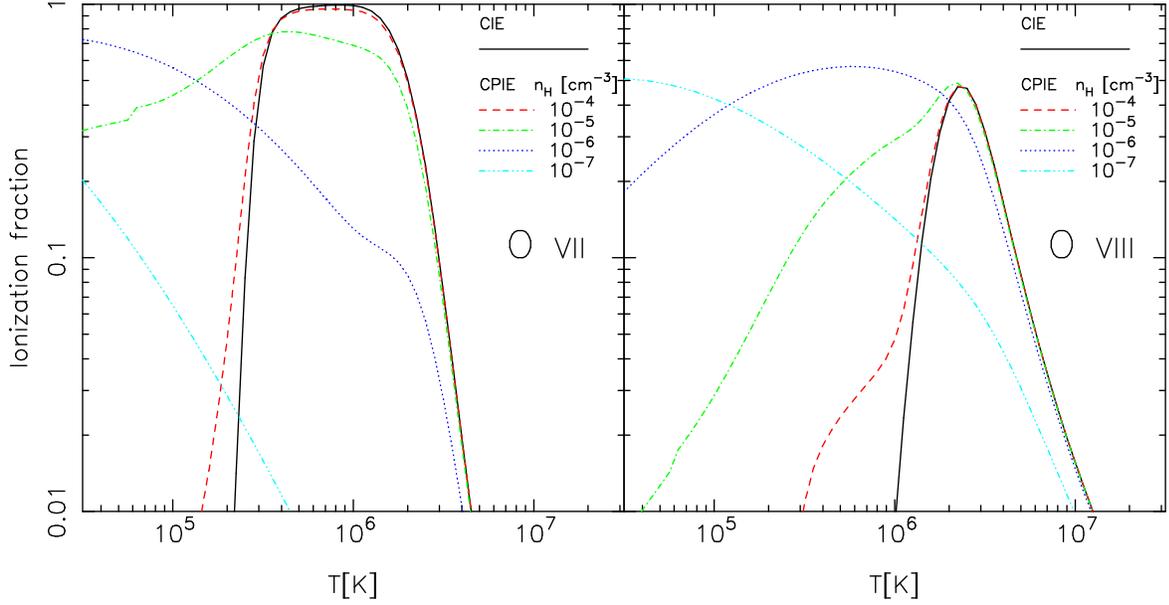

 \begin{center}
 \rotatebox[origin=c]{270}{
 \FigureFile(80mm,80mm){figure1.eps}} \caption{Ionization
 fractions of O{\sc vii} and O{\sc viii} as a function of
 temperature. Solid lines represent results for CIE (Collisional
 Ionization Equilibrium), while other different lines indicate those for
 CPIE (Collisional and Photo-Ionization Equilibrium) adopting the
 hydrogen number density $n_H=10^{-4}$, $10^{-5}$, $10^{-6}$, and
 $10^{-7}$ cm$^{-3}$.  \label{fig:F_oxygen}}
 \end{center}
\end{figure}

Ionization fraction is calculated using a publicly available routine,
SPEX ver. 1.10. Papers I and II have assumed collisional ionization
equilibrium (CIE) for simplicity because detectable emission lines
preferentially come from WHIM with relatively high density. In the
absorption lines, however, lower-density WHIM may become important. So
in this paper we take account of the photo-ionization effect and compute
the ionization fraction of metals under the collisional and
photo-ionization equilibrium (CPIE) \footnote{ Non-equilibrium
ionization effects on WHIMs have recently considered by \citet{CenFang}
and \citet{YoshikawaSasaki}. The latter paper concluded that the
detectability of WHIMs is largely unaffected by the non-equilibrium
effects, although the line ratio of O{\sc{vii}} and O{\sc{viii}} is 
systematically overestimated (see their Figures 14 and 16).}.  To do so,
we modified the original SPEX source code, and added routines to
calculate the photo-ionization effect.  For that purpose, we have to
specify the differential energy distribution of the diffuse
extragalactic background radiation.  We simply adopt the sum of the
cosmic X-ray background and the UV background
\citep{Miyaji1998,Shull1999,Chen2003}:
\begin{equation}
 J(E) = J_{\rm CXB}(E) + J_{\rm UV}(E), \qquad (10 {\rm eV} < E < 10
  {\rm keV}) ,
\end{equation}
where
%%-----------------------------------------------------------------------
\begin{equation}
 J_{\mathrm{CXB}} (E) = J_1 \left( \frac{E}{\mathrm{keV}}\right)^{-0.42},
  \label{eq_cxb}
\end{equation}
%%-----------------------------------------------------------------------
with $J_1 = 6.626 \times 10^{-26} 
\, \mathrm{erg \, cm^{-2} \, s^{-1} \, sr^{-1} \, Hz^{-1}} (= 1.602
\times 10^{-8} \, \mathrm{erg \, cm^{-2} \, s^{-1} \, sr^{-1} \,
keV^{-1}}) $, and 
%%-----------------------------------------------------------------------
\begin{equation}
 J_{\mathrm{UV}} (E) = J_2 \left( \frac{E}{13.6 \mathrm{eV}}\right)^{-1.8},
  \label{eq_uvb}
\end{equation}
%%-----------------------------------------------------------------------
with $J_2 = 2.4 \times 10^{-23} \, \mathrm{erg \, cm^{-2} \, s^{-1} \,
sr^{-1} \, Hz^{-1}} (= 5.9 \times 10^{-6} \, \mathrm{erg \, cm^{-2} \,
s^{-1} \, sr^{-1} \, keV^{-1}})$. 
At $E<0.26$ keV, $J_{\rm UV}$ is the dominant component of
the background radiation,
while $J_{\rm CXB}$ dominates at a larger energy region.
The resulting ionization fractions of
O{\sc vii} and O{\sc viii} for CIE and CPIE are plotted in Figure
\ref{fig:F_oxygen}.

Finally we obtain the optical depth of the ionized metal $A^{n+}$ along
a LOS up to $z=0.3$ as
%%%%%%%%%%%%%%%%%%%%%%%%%%%%%%%%%%%%%%%%%%%%%%%%%%%%%%%%
\begin{equation}
 \tau_{A^{n+}} (E) = 
\int_{z=0}^{z=0.3} n_{A^{n+}}(z) \,  
\sigma_{A^{n+}}(E^\prime) \, \frac{d l}{d z} d z, 
\end{equation}
%%%%%%%%%%%%%%%%%%%%%%%%%%%%%%%%%%%%%%%%%%%%%%%%%%%%%%%%
where the rest-frame frequency $E '$ is given by
%%%%%%%%%%%%%%%%%%%%%%%%%%%%%%%%%%%%%%%%%%%%%%%%%%%%%%%%
\begin{equation}
E^\prime= E (1+z)(1+v_r/c) .
\end{equation}
%%%%%%%%%%%%%%%%%%%%%%%%%%%%%%%%%%%%%%%%%%%%%%%%%%%%%%%%
Among hundreds of resonance absorption lines for $ 300 {\rm
eV} < E < 2 {\rm keV}$ \citep{Verner1996}, we choose 
10 lines for 9
ions that are relevant in the WHIM regime (see Table
\ref{tab:abs_species}).  The absorption cross section $\sigma_{A^{n+}}
(E ')$ at $E '$ is computed assuming the Doppler broadening:
%%-----------------------------------------------------------------------
\begin{eqnarray}
 \sigma_{A^{n+}} (E ') = \sum_k \frac{\pi e^2}{m_e c} f_{k,A^{n+}}
  \frac{2\pi \hbar c}{\sqrt{\pi} E_k b_T}
  \exp{\left[ - \frac{c^2}{b_T^2} \frac{(E '-	E_k)^2}{E_k^2}\right]}, 
\end{eqnarray}
%%-----------------------------------------------------------------------
where $k$ denotes the index of different absorption lines of $A^{n+}$,
$f_{k,A^{n+}}$ and $E_k$ are the oscillator strength and the energy of
the $k$-th line, and $b_T= (2 k_B T/m_A)^{1/2}$ is the Doppler
$b$-parameter ($m_A$ is the mass of the metal $A$).

Here we consider the thermal broadening only.  Since the
natural widths of the lines of our interest are less than 0.02eV, we can
safely neglect them and thus use Gaussian rather than the Voigt profile.
Strictly speaking, the internal turbulence is expected to be partly
incorporated in our simulation data, but may not be accurate enough.
Because we are mainly interested in unsaturated lines, the additional
effect of the turbulence is not supposed to be large.  Thus we decided
to neglect it in the current analysis.

\subsection{Mock Absorption Spectra of WHIM toward a bright QSO} 

Now we are in a position to compute mock absorption spectra of WHIM
toward a bright QSO.

\begin{figure}[htbp]
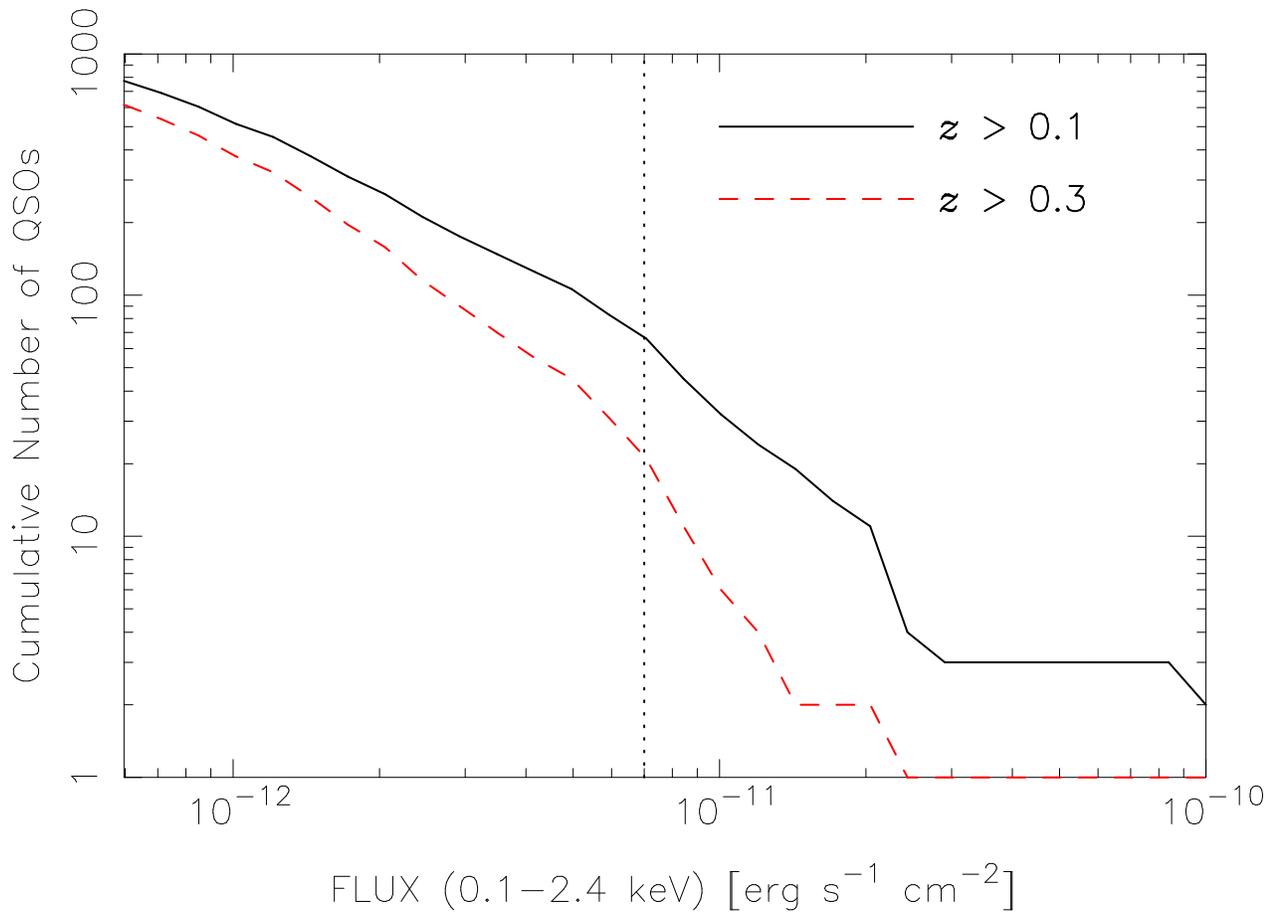

 \begin{center}
 \rotatebox[origin=c]{270}{
  \FigureFile(120mm,120mm){figure2.eps}}
 \end{center}
 \caption{ Cumulative number counts of QSOs in X-ray from
ROSAT data as a function of their 0.1 -- 2.4 keV flux. Solid and dashed
lines indicate QSOs with $z>0.1$ and $z>0.3$, respectively. 
\label{fig:qsoflux}}
\end{figure}

Figure \ref{fig:qsoflux} shows the cumulative number counts of QSOs in
X-ray from ROSAT data \citep{Brinkmann1997, Yuan1998}.  
We adopt the flux of $f(0.1-2.4 ~{\rm keV})=7 \times 10^{-12}
~{\rm erg~s^{-1} cm^{-2}}$ as our fiducial value.
This value is marked in the vertical dotted line.
Beyond the value, we find more than 20 QSOs with $z>0.3$. 
For definiteness, we adopt the power-law spectrum as
%%%%%%%%%%%%%%%%%%%%%%%%%%%%%%%%%%%%%%%%%%%%%%%%%%%%%%%%%%%%%%%%%%%%%%%%
\begin{equation}
 F_{\mathrm{QSO}}(E) = 1.4 \times 10^{-12}
  \left( \frac{E}{1 {\rm keV}} \right)^{-1.5}
  ~{\rm erg~s^{-1} cm^{-2} keV^{-1}}. 
\end{equation}
%%%%%%%%%%%%%%%%%%%%%%%%%%%%%%%%%%%%%%%%%%%%%%%%%%%%%%%%%%%%%%%%%%%%%%%%

We consider mock observations of a QSO with 30 ksec exposure.
Then fluence ${\cal F}$
is computed by taking account of the absorption due to Galaxy and to
the intervening WHIM:
%%%%%%%%%%%%%%%%%%%%%%%%%%%%%%%%%%%%%%%%%%%%%%%%%%%%%%%%%%%%%%%%
\begin{eqnarray}
\label{eq:fE}
 {\cal F}(E) &=& {\cal F}_0 (E) 
  \exp{\left[- \sum_{A^{n+}} \tau_{A^{n+}} (E) \right]} 
~~ \mathrm{[erg/cm^2/eV]},  \\ 
\label{eq:fE0}
{\cal F}_0(E) &=& 
\exp{\left[-\sigma_{gal}(E) N_{H}\right]} \int_{t_i}^{t_f}
 F_{\mathrm{QSO}}(E) dt ~~ \mathrm{[erg/cm^2/eV]},
\end{eqnarray}
%%%%%%%%%%%%%%%%%%%%%%%%%%%%%%%%%%%%%%%%%%%%%%%%%%%%%%%%%%%%%%%%
where 
${\cal F}_0(E)$ determines the observable continuum level free 
from the WHIM absorption.  The Galactic absorption is computed using the
absorption cross section $\sigma_{gal}(E)$ of \citet{Morrison1983} who
adopt 1 solar metallicity and the neutral hydrogen column density
$N_{H} = 3 \times 10^{20} \mathrm{cm^{-2}}$.
The resulting Galactic extinction factor, 
$\exp[-\sigma_{gal}(E) N_{H}]$, is
$\approx 0.80$ at $E=0.6$keV.
For simplicity, we neglect the additional intrinsic absorption.

\subsection{Possible application to bright GRB afterglows}

The same methodology can be applied to bright GRB afterglows.
In addition to its intrinsic brightness, the more important advantage of
GRB for the WHIM science lies in its transient nature; once absorption
lines due to WHIM are detected, one can perform deep follow-up
observations to look for the emission counterpart (Papers I and II)
after the GRB is completely faded.  
Successful detections of both absorption and emission features will
offer unambiguous evidence of WHIM, although not so easy in reality.

\citet{Piro2004} pointed out that the
average spectrum energy distribution of a typical GRB afterglow is
approximated as
%%%%%%%%%%%%%%%%%%%%%%%%%%%%%%%%%%%%%%%%%%%%%%%%%%%%%%%%%%%%%%%%%%%%%%%%
\begin{eqnarray}
 F_{\mathrm{GRB}}(t,E) &=& F_0 \left(\frac{t}{40000 {\rm s}} \right)^{-
\delta}  
\left( \frac{E}{1 {\rm keV}} \right)^{- \alpha}, \\
 \delta &=& 1.2 \pm 0.2,~~ \alpha = 1.13 \pm 0.07.
\end{eqnarray}
%%%%%%%%%%%%%%%%%%%%%%%%%%%%%%%%%%%%%%%%%%%%%%%%%%%%%%%%%%%%%%%%%%%%%%%%
For six years between 1997 and 2002, BeppoSax observed 8\% of the
entire sky \citep{Jager1997}.  The brightest GRB
afterglow detected has $F_0 = 6 \times 10^{-12} ~ \mathrm{erg~ s^{-1}
cm^{-2} keV^{-1}}$ approximately \citep{Piro2004,DePasquale}. 

We consider mock observations of a GRB afterglow for $t_i < t < t_f$
after the GRB event ($t=0$). 
If we observe this GRB afterglow with $(t_i, t_f)=(1 {\rm day}, 1~{\rm
day}+30 {\rm ksec})$, the fluence is comparable with one for our fiducial
QSO. 
For definiteness, we adopt the flux as our fiducial value:
%%%%%%%%%%%%%%%%%%%%%%%%%%%%%%%%%%%%%%%%%%%%%%%%%%%%%%%%%%%%%%%%%%%%%%%%
\begin{equation} \label{eq_GRB} F_{\mathrm{GRB}}(t,E) = 6 \times
10^{-12} \left( \frac{t}{40000{\rm s}}\right)^{-1.2} \left(
\frac{E}{1{\rm keV}} \right) ^{-1.13} \mathrm{erg~s^{-1} cm^{-2}
keV^{-1}}.  \end{equation}
%%%%%%%%%%%%%%%%%%%%%%%%%%%%%%%%%%%%%%%%%%%%%%%%%%%%%%%%%%%%%%%%%%%%%%%%
Assuming a nominal observational efficiency of 50\%, GRB afterglows
brighter than our fiducial value (Eq.[\ref{eq_GRB}]) are roughly
expected to be $\approx 1/0.5/0.08/6 = 4$ per year over the entire
sky. In reality, however, the above estimate is statistically limited
so far, and a better estimate will be given from the on-going {\it
Swift} observation.

\section{Results}

\subsection{Identification of absorption lines in mock spectra}

\begin{figure}[htbp]
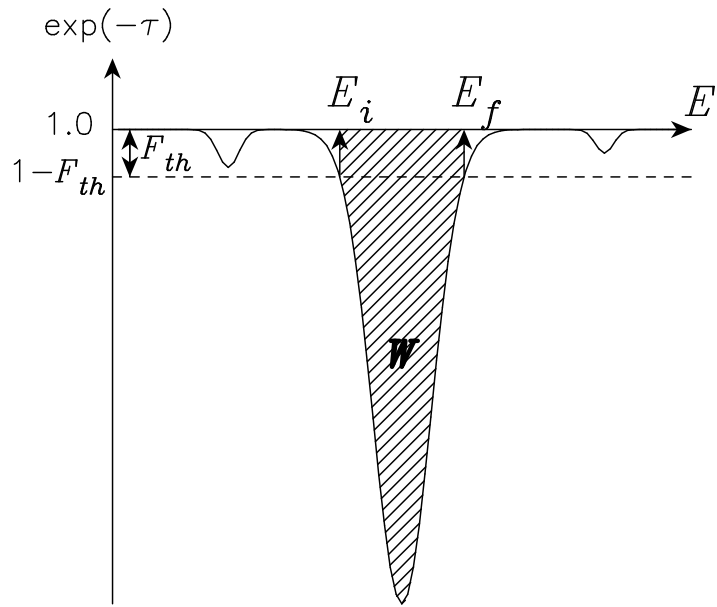

 \begin{center}
 \rotatebox[origin=c]{270}{
 \FigureFile(80mm,80mm){figure3.eps}}
 \caption{Schematic identification method of absorption lines. 
 \label{fig:identify_line}}
 \end{center}
\end{figure}

Once mock spectra are constructed, our next task is to attempt the mock
observation, and in particular, to identify absorption lines in an
objective manner. For this purpose, we follow the procedure described in
\citet{Fang2002a}. Consider a schematic example of normalized absorption
line profiles, $\exp{(-\tau)}$ as in Figure \ref{fig:identify_line}.  We
set a threshold value $F_{\mathrm{th}} = 10^{-5}$, and identify a
contiguous region below the threshold $1-F_{\mathrm{th}}$ as an
absorption line system. Suppose the line profile which first
down-crosses the $1-F_{\mathrm{th}}$ at $E=E_i$ and then up-crosses at
$E=E_f$. The equivalent width $W$ (in the observer's frame) of this
system centered at $E=E_l$ is calculated by integrating
$1-\exp{(-\tau)}$ over $E_i<E<E_f$:
%%%%%%%%%%%%%%%%%%%%%%%%%%%%%%%%%%%%%%%%%%%%%%%%%%%%%%%%%%%%%%%%%%%%%
\begin{equation}
W_l(E_l) = \int_{E_i}^{E_f} \left[1-e^{-\tau(E)}\right] dE .
\label{eq:EW}
\end{equation}
%%%%%%%%%%%%%%%%%%%%%%%%%%%%%%%%%%%%%%%%%%%%%%%%%%%%%%%%%%%%%%%%%%%%%
If the line system is identified as a particular species with the
oscillator strength $f_l$, the equivalent width in its unsaturated
regime is rewritten in terms of the corresponding column density $N_l$
as:
%%%%%%%%%%%%%%%%%%%%%%%%%%%%%%%%%%%%%%%%%%%%%%%%%%%%%%%%%%%%%%%%%%%%%
\begin{equation}
N_l = \frac{9\times 10^{14}}{f_l} \frac{W_l}{0.1 {\rm eV}}~[{\rm cm}^{-2}] .
\label{eq:column-dens}
\end{equation}
%%%%%%%%%%%%%%%%%%%%%%%%%%%%%%%%%%%%%%%%%%%%%%%%%%%%%%%%%%%%%%%%%%%%%

In the present case, the $S/N$ is determined by the energy resolution
of the detector (for {\it XEUS}, the effective area $S_{\rm eff} =
60000 {\rm cm^2}$ and the energy resolution $\Delta E=2 {\rm eV}$; for
comparison, intrinsic widths of the lines of our interest are typically
below 1 eV).  Therefore the $S/N$ at the energy of the line center,
$E_l$, 
is estimated as
%%%%%%%%%%%%%%%%%%%%%%%%%%%%%%%%%%%%%%%%%%%%%%%%%%%%%%%%%%%%%%%%%%%%%
\begin{equation}
S/N(E_l) = \frac{[{\cal N}_0(E_l) - {\cal N}(E_l)] \Delta E}
{\sqrt{{\cal N}(E_l) \Delta E}}
\label{eq_sn}
\end{equation}
%%%%%%%%%%%%%%%%%%%%%%%%%%%%%%%%%%%%%%%%%%%%%%%%%%%%%%%%%%%%%%%%%%%%%
simply from the Poisson statistics of the observed photons, where
${\cal{N}}$ and ${\cal{N}}_0$ are the photon number 
counts [ph/eV] with absorption and without absorption (continuum level),
respectively. 

\begin{figure}[htbp]
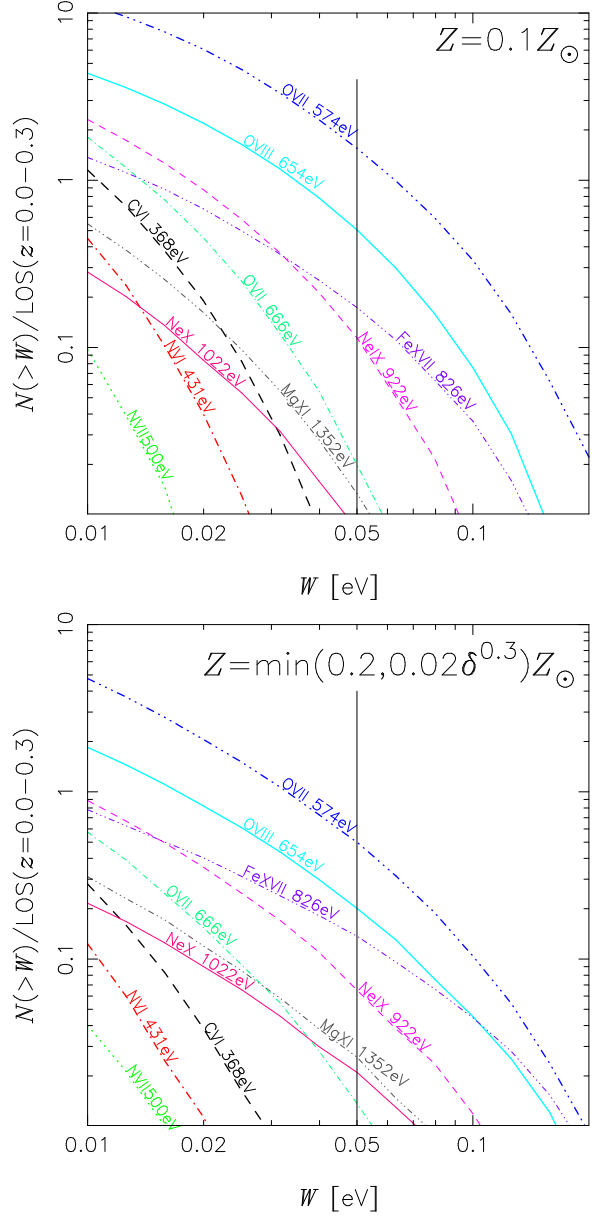

 \begin{center}
 \rotatebox[origin=c]{270}{
 \FigureFile(80mm,80mm){figure4a.eps}
 \FigureFile(80mm,80mm){figure4b.eps}}
 \end{center}
 \caption{Cumulative distribution of expected number of absorption line
systems for $0<z<0.3$ as a function of their equivalent width; {\it
Upper panel:} $Z=0.1Z_\odot$, {\it Lower panel:} $Z/Z_\odot=\mathrm{min}
[0.2 , 0.02(\rho/\bar{\rho}_{\mathrm{b}})^{0.3}]$.  Solid
vertical lines indicate our fiducial detection limit for the equivalent
width ($W=0.05$eV).  \label{fig:cumulative_ew}}
\end{figure}

\begin{figure}[htbp]
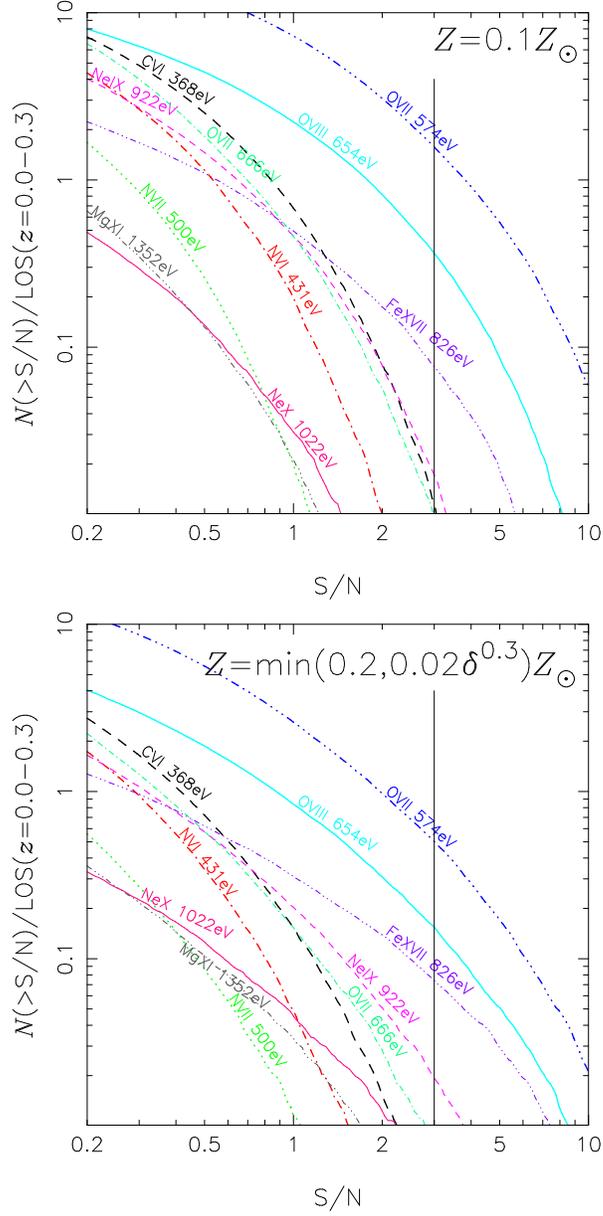

 \begin{center}
 \rotatebox[origin=c]{270}{
  \FigureFile(80mm,80mm){figure5a.eps}
  \FigureFile(80mm,80mm){figure5b.eps}}
 \end{center}
 \caption{Cumulative distribution of expected number of absorption line
systems for $0<z<0.3$ as a function of their detection signal-to-noise
ratios. The plots are converted from those of
Fig. \ref{fig:cumulative_ew} assuming the {\it XEUS} observation toward
a QSO discussed in \S 2.3. 
Solid vertical lines indicate our fiducial detection limit ($S/N=3$).
 \label{fig:cumulative_sn}}
\end{figure}

We compute the equivalent width and the $S/N$ of absorption line systems 
for 6400 random LOSs in the $5^\circ \times 5^\circ$ region assuming the
fluence described in \S 2.3.  Figure \ref{fig:cumulative_ew}
plots the cumulative number distribution of the equivalent width of
absorption line systems per LOS ($z=0.0-0.3$) from our mock observation
for 10 strong lines (Table \ref{tab:abs_species}).  On average, O {\sc
vii} (574 eV) is the 
most prominent line in terms of the equivalent width, and O {\sc viii}
(654 eV) and Fe {\sc xvii} (826 eV) are the next.  Figure
\ref{fig:cumulative_sn} shows the corresponding cumulative number
distribution of $S/N$. Note that the value of S/N should be replaced by 
%%%%%%%%%%%%%%%%%%%%%%%%%%%%%%%%%%%%%%%%%%%%%%%%%%%%%%%%%%%%%%%%%%%%%%%
\begin{eqnarray}
S/N &=& (S/N)_{\rm fiducial}
\left(
\frac{F_0}{6 \times 10^{-12} ~ \mathrm{[erg~ s^{-1} cm^{-2}keV^{-1}}]}
\right)^{1/2} \cr
&=& (S/N)_{\rm fiducial}
\left(\frac{{\cal N}_0(E=\mathrm{500eV})}{8800 \, \mathrm{[ph/eV]}}
\right)^{1/2} 
\end{eqnarray}
%%%%%%%%%%%%%%%%%%%%%%%%%%%%%%%%%%%%%%%%%%%%%%%%%%%%%%%%%%%%%%%%%%%%%%%
if $F_0$, and more generally the number of photons during the exposure
time, ${\cal N}_0$, are different from our fiducial values.  In our
adopted model for the QSO and the {\it XEUS} performance, O
{\sc vii} 574 eV and O{\sc viii} 654 eV lines achieve $S/N = 3$ when
their equivalent widths are $W=0.05$ eV. We take this value as our
detection limit. 

We find that the number of absorption systems which have $S/N \ge 3$ is
1.58/LOS (O{\sc vii} 574 eV) and 0.37/LOS (O {\sc viii} 654 eV) as
summarized in Table \ref{tab:abs_oxy}. The other eight lines exhibit
below 0.1 systems/LOS. The number of the case that both O{\sc vii} (574
eV) and O{\sc viii} (654 eV) exhibit $S/N \ge 3$ at the same position of
WHIM is 0.35/LOS.

In order to check the dependence on the assumed metallicity model, we
also consider two other metallicity models. One is a phenomenological
model adopted in Paper II (the radiation pressure ejection model of
\cite{Aguirre2001}):
%%-----------------------------------------------------------------------
\begin{equation}
\label{eq:Z_aguirre}
   Z_{A,j}/Z_{A,\odot}=\mathrm{min} 
\left[0.2 , 0.02(\rho_j/\bar{\rho}_{\mathrm{b}})^{0.3}\right] ,
\end{equation}
%%-----------------------------------------------------------------------
where $\bar{\rho}_{\rm b}$ is the mean baryon density at redshift
$z$. The other adopts a constant metallicity comparable to the typical
value for intra-cluster medium:
%%-----------------------------------------------------------------------
\begin{equation}
   Z_{A,j}=0.3 Z_{A,\odot}.
\end{equation}
%%-----------------------------------------------------------------------
Figure \ref{fig:metal_dep} shows the dependence of the cumulative number
distribution of oxygen absorption line systems on the metallicity
models; O {\sc vii} (574eV) and O {\sc viii} (654eV) in thick and thin
lines, respectively. Note that $Z_{A,j}>0.1 Z_{A,\odot}$ is satisfied
only in relatively dense regions $\rho_j/\bar{\rho}_{\mathrm{b}}>214$
(eq.[\ref{eq:Z_aguirre}]).  This is why our fiducial model ($Z_{A,j} =
0.1 Z_{A,\odot}$) has more absorption line systems than the model of
equation (\ref{eq:Z_aguirre}).

We also show the results toward our fiducial GRB afterglow in Table 3.
These are almost equivalent to those toward our fiducial QSO (Table 2).
Thus, we concentrate on the results toward QSOs in what follows.

\subsection{Detectability of Oxygen Absorption Line Systems}

Since the most prominent absorption lines detectable by {\it XEUS} are
O{\sc vii} and O{\sc viii}, we focus on these lines in what follows.
Figure \ref{fig:columndens_map} shows column density maps of O{\sc vii}
({\it Left panel}) and O{\sc viii} ({\it Right panel}) for $0<z<0.3$.
As indicated in those panels, we select three different LOSs (A, B, and
C). The mock spectra for those LOSs are shown in Figure
\ref{fig:mock_spectra}; spectrum A exhibits many prominent absorption
lines, and spectrum B has one strong line system while spectrum C has
none.

We find two distinct absorption line systems which are located at
$z=0.12$ (marked by $\dagger$) and $z=0.29$ (marked by $\ddagger$) along
the LOS of A; O{\sc vii} ($z=0.29$, $E=445$eV, $W=0.16$eV), O{\sc viii}
($z=0.29$, $E=507$eV,$W=0.09$eV), O{\sc vii} ($z=0.12$, $E=514$eV,
$W=0.10$eV), and O{\sc viii} ($z=0.12$, $E=585$eV, $W=0.12$eV).  Figure
\ref{fig:map12} plots the column density map of O{\sc vii} ({\it Left
panels}) and O{\sc viii} ({\it Right panels}) but for $0.11<z<0.14$
({\it Upper panels}) and for $0.26<z<0.30$ ({\it Lower panels}) so as to
depict the WHIM responsible for the absorption systems in the spectrum
A.

Figure \ref{fig:t_delta} shows the mass-weighted (within each cell)
temperature and the over-density distributions along the three LOSs. The
top two panels for the region A have high over-density (10-100) and high
temperature ($10^6-10^7 \mathrm{K}$) WHIM clumps at $z=0.12$ and
$z=0.29$. They are indeed responsible for the prominent lines in the
spectrum A of Figure \ref{fig:mock_spectra}.  Spectrum B has one
absorption system at $z=0.22$ (O{\sc vii} 574 eV $W=0.06$eV) as
indicated in Figures \ref{fig:map12} and \ref{fig:t_delta}. In reality,
however, the identification of the line redshift is very difficult from
a single line alone. One may assume statistically that the strongest
line in a single LOS should be that of O{\sc vii} 574 eV, and then infer
the redshift $574/470-1=0.22$. In this particular example, it turns out
a correct guess, but may not work always.  Finally spectrum C has no
prominent absorption feature. Indeed Figure \ref{fig:t_delta} shows that
there is no region with over-density $\delta \equiv \rho/\bar{\rho} >10$
along the LOS.

Figure \ref{fig:rho_T} plots the distributions of mass-weighted
temperatures and over-densities of WHIM averaged over each absorption
line system of O{\sc vii} (574 eV) and O {\sc viii} (654 eV).
Spatial boundaries of absorption line systems are somewhat
ambiguous. We define here the cosmic mean density as the threshold as
illustrated in Figure \ref{fig:identify_whim}, and we identify one
connected region exceeding $\delta=1$ as a single absorption system even
if there are many density peaks and valleys inside.
Then we compute its temperature and over-density by
mass-weighted averaging over the relevant SPH particles within the
entire line system.

The different symbols indicate the range of the equivalent width $W$ of
the absorption line systems; circles, crosses and triangles correspond
to the systems with $W \ge 0.12$ eV, $0.07 \le W \le 0.12$ eV, and $0.05
\le W \le 0.07$ eV, respectively.  Both O{\sc vii} and O {\sc viii} can
probe WHIM with $T > 10^6 - 10^7 $ K and over-density of $\delta =10$ to
1000. WHIM with $T<10^6$ K, however, can be traced only by O {\sc vii}.

Figure \ref{fig:r_h_w} plots the fraction of gas contained in oxygen
absorption line systems (O {\sc vii} in
{\it Left panels} and O {\sc viii} in
{\it Right panels}) which have the equivalent width larger
than the specified value $W$.  In practice we first select all oxygen
absorption line systems along the $i$-th LOS whose equivalent width
exceeds $W$, and sum up their hydrogen gas column densities. We denote
the sum as $N_{H,i}^{\mathrm{sim}}(>W)$. Then we take the following
ratio averaged over 6400 LOSs:
%%%%%%%%%%%%%%%%%%%%%%%%%%%%%%%%%%%%%%%%%%%%%%%%%%%%%%%%%%%%%%%%%%
\begin{equation}
\label{eq:r_h_w}
R_H(>W) = \frac{\displaystyle \sum_{i=1}^{6400}N_{H,i}^{\mathrm{sim}}(>W)}
{\displaystyle \sum_{i=1}^{6400}N_{H,i}^{\mathrm{sim}}(>0)}
\end{equation}
%%%%%%%%%%%%%%%%%%%%%%%%%%%%%%%%%%%%%%%%%%%%%%%%%%%%%%%%%%%%%%%%%%
for O {\sc vii} and O {\sc viii} separately.  In reality this is an {\it
unobservable} quantity, but provides a {\it simulated} gas mass fraction of
oxygen absorption line systems which have a detectable line width. More
importantly, this is the essential factor in properly estimating the
total cosmic baryon density from the detected fraction of oxygens (\S
4.2).

In the upper panels, three shaded regions correspond to the temperature
ranges $T>10^7$ K, 
$10^6<T<10^7$ K, and $T<10^6$ K from top to bottom, and in the lower
panels, three shaded regions correspond 
to the density contrast ranges $\delta>100, 10 < \delta < 100, \delta <
10$ from top to bottom.
Comparison of the left and right panels suggests that O{\sc viii} traces 
higher temperature and denser regions than O {\sc vii}. In
the case of O {\sc vii} (574 eV), we can detect only 32 \% of baryon
above the detection limit ($W \sim 0.05$ eV; \S 3.1). The region of
$T>10^7$ K accounts for 19 \% of detectable baryons, which may be
located in virialized regions like intra-cluster medium and thus may be
already detected through their thermal bremsstrahlung. The regions of
$10^6<T<10^7$ K and $T<10^6$ K account for 68 \% and 12 \% of detectable
baryons, which can be regarded as WHIM. Note that one can detect the
absorption lines of O {\sc vii} (574eV) with $T<10^6$ K, while it is
almost impossible to find their emission counterparts (Paper I).

\section{Physical Properties of the identified WHIM}

\subsection{O{\sc vii} and O{\sc viii} Absorption Lines}

As we have seen, the most prominent absorption lines are O{\sc vii} (574
eV) and O{\sc viii} (654 eV). On average, 
35 \% of random LOSs have an
intervening clump of WHIM which exhibits both absorption lines with $S/N
\ge 3$.  In this case, the ratio of their equivalent widths should
provide the temperature of the WHIM {\it in principle}. Consider a clump
of WHIM which has a uniform temperature $T$ and density $\rho$, then the
ratio of the equivalent widths for unsaturated lines that
are relevant here is written as
%%%%%%%%%%%%%%%%%%%%%%%%%%%%%%%%%%%%%%%%%%%%%%%%%%%%%%%%%%%%%%%%%%%%%%
\begin{equation}
  \frac{W_{\mathrm{O VII, 574 eV}}}{W_{\mathrm{O VIII, 654 eV}}} \sim
   \frac{f_{\mathrm{O VII, 574 eV}} F_{\mathrm{O VII}} (T,\rho)}{
   f_{\mathrm{O VIII, 654 eV}} F_{\mathrm{O VIII}} (T,\rho)},
\label{eq_t_relation}
\end{equation}
%%%%%%%%%%%%%%%%%%%%%%%%%%%%%%%%%%%%%%%%%%%%%%%%%%%%%%%%%%%%%%%%%%%%%%
where $f$ is the oscillator strength and $F$ is the ionization fraction.
The WHIM which exhibits O{\sc viii} (654 eV) absorption has temperature
of $10^6 - 10^7$ K (Figure \ref{fig:rho_T}). In this temperature range, the
ionization fraction approaches the value in CIE, and the ratio of the
oxygen equivalent widths, eq.(\ref{eq_t_relation}), is expected to be
insensitive to the density, $\rho$.

Figure \ref{fig:scat_temp} provides the scatter plot of the ratio of O
{\sc vii} and O {\sc viii} equivalent width versus the mass-weighted
average temperature for systems where which both O {\sc vii} (574 eV)
and O {\sc viii} (654 eV) exhibit $S/N \ge 3$. Three solid lines
indicate the theoretical curves in Eq. (\ref{eq_t_relation}) applying
different hydrogen number density $n_H = 10^{-5}$, $10^{-6}$, and
$10^{-7}$ $\mathrm{cm^{-3}}$ from top to bottom. Inhomogeneous density
and temperature structures are responsible for the departure from the
theoretical model which assumes uniform distribution.  Figure
\ref{fig:prof} shows an example of sub-structure in a simulated
absorption line system. Temperature is the highest near the center of
the clump, and over-density is also high. Therefore, ionization
fractions of O{\sc vii} and O{\sc viii} are rather suppressed.  This
anti-correlation of oxygen abundances inside the inhomogeneous
distribution of WHIMs explains why the simulated systems in Figure
\ref{fig:scat_temp} preferentially lie at higher temperature regime than
the theoretical prediction.

\subsection{Estimating $\Omega_b$ from WHIM observations}

If absorption lines of WHIM are detected along a particular line of
sight, one would naturally attempt to estimate the contribution of WHIM
to the total cosmic baryon density,
$\Omega^{\mathrm{WHIM}}_b$. \citet{Nicastro2005} indeed reported
$\Omega_{\rm b}^{\rm WHIM} = (2.7 {+3.8 \atop -1.9}\%) \times 10^{-[{\rm
O}/{\rm H}]_{-1}}$ using two O{\sc vii} absorption lines.
We perform statistical analysis of
$\Omega^{\mathrm{WHIM}}_b$ estimation using our mock data.

Let us first define the estimated fraction of WHIM with signal-to-noise
ratio exceeding a given threshold, $S/N$, along one LOS as follows:
%%%%%%%%%%%%%%%%%%%%%%%%%%%%%%%%%%%%%%%%%%%%%%%%%%%%%%%%%%%%%%%%%%%%%%
\begin{eqnarray}
\label{eq:omega_gas_est}
 \Omega_{\rm gas}^{\mathrm{WHIM, est}}(>S/N) 
&=& \frac{1}{d_c(z=0.3)\rho_c}  
\sum_i  \frac{N_{H,i}^{\mathrm{est}}(>S/N)}{X (1+z_i)^2}, \\
\label{eq:nh_est}
 N_{H,i}^{\mathrm{est}}(>S/N)
&=& \frac{N_{\mathrm{OVII},i}(>S/N)}{F_{\mathrm{OVII},i}(T)}
\left(\frac{n_{H}}{n_{O}}\right)
\end{eqnarray}
%%%%%%%%%%%%%%%%%%%%%%%%%%%%%%%%%%%%%%%%%%%%%%%%%%%%%%%%%%%%%%%%%%%%%%
where $i$ denotes an index for absorption line systems, $X(=0.755)$ is
the hydrogen mass fraction, and $d_c(z=0.3)$ is the comoving
distance from $z=0$ to $z=0.3$. We convert the observed
equivalent width into $N_{\mathrm{OVII}}(>S/N)$ using
Eq. (\ref{eq:column-dens}) assuming that the line is unsaturated, i.e.,
in its linear regime of the growth curve. 
At present, no simultaneous detection of O{\sc vii} and
O{\sc viii} absorption lines in WHIM is reported. Consider first,
however, the case in which both O {\sc vii} and O {\sc viii} are
detected.
Then we estimate temperature $T$ using Eq. (\ref{eq_t_relation}) and derive
the ionization fraction $F_{\mathrm{OVII}} (T)$ assuming collisional
ionization equilibrium. If we multiply the hydrogen-to-oxygen number
ratio, we have an estimate of the hydrogen column density for the $i$-th
absorption line system, $N_{H,i}^{\mathrm{est}}(>S/N)$. In practice, we
adopt $(n_{O}/n_H) = 0.1 (n_{O}/n_H)_\odot = 10^{-4.07}$
\citep{Anders1989}. Finally this is converted to the density parameter of
gas, $\Omega_{\rm gas}^{\mathrm{WHIM, est}}(>S/N)$ responsible for
absorption exceeding the detection $S/N$.

As we have seen in the previous sections, the inhomogeneity of the
temperature and density within one absorption system would result in a
systematic underestimate of $N_{H,i}^{\mathrm{est}}(>S/N)$. In order to
check the systematic effect as well as the assumption of CIE in equation
(\ref{eq:nh_est}) (while we did adopt CPIE in creating mock spectra from
simulation data), we compare $N_{H,i}^{\mathrm{est}}(>S/N)$ with
$N_{H,i}^{\mathrm{sim}}(>S/N)$ directly computed from simulation data
(\S 3.2). Figure \ref{fig:nhest-nhsim} clearly shows the degree of the
underestimate bias.

In order to estimate the total baryon density, $\Omega_{\rm b}^{\rm
est}$, from $\Omega_{\rm gas}^{\mathrm{WHIM, est}}(>S/N)$, we have to
correct for (i) the fraction of gas below the detection limit, (ii) the
fraction of stars in galaxies, and (iii) the above systematic
bias. Therefore we write the estimate as
%%%%%%%%%%%%%%%%%%%%%%%%%%%%%%%%%%%%%%%%%%%%%%%%%%%%%%%%%%%%%%%%%%%%%%
\begin{eqnarray}
\label{eq:omega_est}
 \Omega_{\rm b}^{\rm est}
= \frac{\Omega_{\rm gas}^{\mathrm{WHIM, est}}(>S/N)}{R_H(>S/N)}
\frac{f_{\rm gas}+f_{\rm star}}{f_{\rm gas}}
\Big\langle \frac{N_{H}^{\mathrm{sim}}(>S/N)}
{N_{H}^{\mathrm{est}}(>S/N)}\Big\rangle .
\end{eqnarray}
%%%%%%%%%%%%%%%%%%%%%%%%%%%%%%%%%%%%%%%%%%%%%%%%%%%%%%%%%%%%%%%%%%%%%%
In practice, we choose $S/N=3$ in the following.  We compute the
fraction of WHIMs which exhibit both O{\sc vii} and O{\sc viii}
absorptions (similarly as in Figure \ref{fig:r_h_w}), and find
$R_H(>3\sigma)=0.22$. As for the second factor, we evaluate directly
from simulations; $f_{\rm gas}=0.79$ and $f_{\rm star}=0.21$.  As Figure
\ref{fig:nhest-nhsim} indicates, the last correction factor
significantly varies from line to line, and it is not easy to choose a
relevant average value. If we adopt the mean value as plotted in
triangles, $\langle N_{H}^{\mathrm{sim}}(>3 \sigma)/
N_{H}^{\mathrm{est}}(>3 \sigma)\rangle \sim 1/4$. Adopting those
correction factors, we plot the histogram of $\Omega_{\rm b}^{\rm est}$
in Figure \ref{fig:omega}.

{\it Upper panel}~ is the case where both O{\sc vii} and O{\sc viii}
absorptions are detected (1903 LOSs out of 6400 LOSs).
The mean value of
the resulting $\Omega_{\rm b}^{\rm est}$ averaged over the entire LOSs
(even including the non-detected 4097 LOSs) amounts to $0.023$ (plotted
in the dotted line). This is in reasonable agreement with the value
adopted in the simulation, 0.03.  On the other hand, it clearly
demonstrates the fact that any estimate based on the positive detection
alone may significantly overestimate the real value; it is essential to
take account of the above correction factors properly.

{\it Lower panel} ~in Figure \ref{fig:omega} indicates the 5217 LOSs
along which only O {\sc vii} absorption lines are detected.  In this
case, one has to assume the temperature {\it a priori}, and we show
results for $T=10^{5.5}$, $10^{6.0}$, and $10^{6.5}$ K.  The temperature
dependence is very strong and even not monotonic.  This reflects the
strong dependence of O{\sc vii} fraction on temperature as exhibited in
Figure \ref{fig:F_oxygen}. The two peaks in the histogram correspond to
LOSs which have single and double absorption line systems.
If we assume $T=10^6$ K (the ionization fraction of O{\sc
vii} peaks around this temperature),  O{\sc vii} absorption lines alone
provides a fairly accurate estimate for $\Omega_{\rm b}$
\citep{Nicastro2005}.

\section{Conclusions and Discussion}

In this paper, we have examined the detectability of WHIM via the
absorption lines in a bright QSO spectra using the
cosmological hydrodynamic simulation. We created the mock {\it XEUS}
absorption spectra toward a bright QSO assuming the
collisional and photo-ionization equilibrium.  We conclude that on
average {\it XEUS} will detect 1.58 (O {\sc vii} 574 eV) and 0.37
(O{\sc viii} 654 eV) per a random line-of-sight up to $z=0.3$ ($S/N \ge 
3$).

If both O{\sc vii} and O{\sc viii} absorptions are detected ($\sim 40\%$
chance) for the same WHIM clump, one can estimate its temperature from
their line ratios and attempt to infer the cosmic baryon density. As we
discussed in detail, however, the reliable estimate requires several
careful correction factors and still the resulting distribution function
of the estimate is fairly wide. So any attempt on the basis of small
statistical samples should be interpreted with caution.

\begin{figure}[htbp]
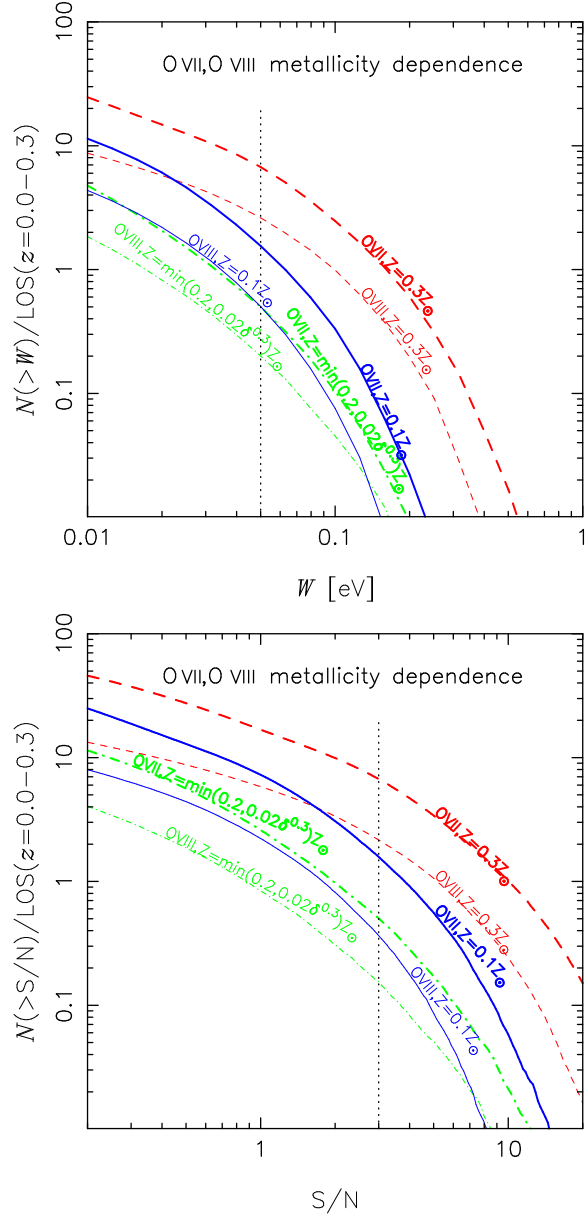

 \begin{center}
 \rotatebox[origin=c]{270}{
  \FigureFile(80mm,80mm){figure6a.eps}
  \FigureFile(80mm,80mm){figure6b.eps}}
 \end{center}
 \caption{ Cumulative distribution of O{\sc vii} and O{\sc viii}
absorption lines for different metallicity models as a function of $W$
({\it Upper panel}) and $S/N$ ({\it Lower panel}). Dotted 
vertical lines indicate
our fiducial detection limits ($W=0.05$eV and $S/N=3$).
\label{fig:metal_dep}}
\end{figure}

\begin{figure}[htbp]
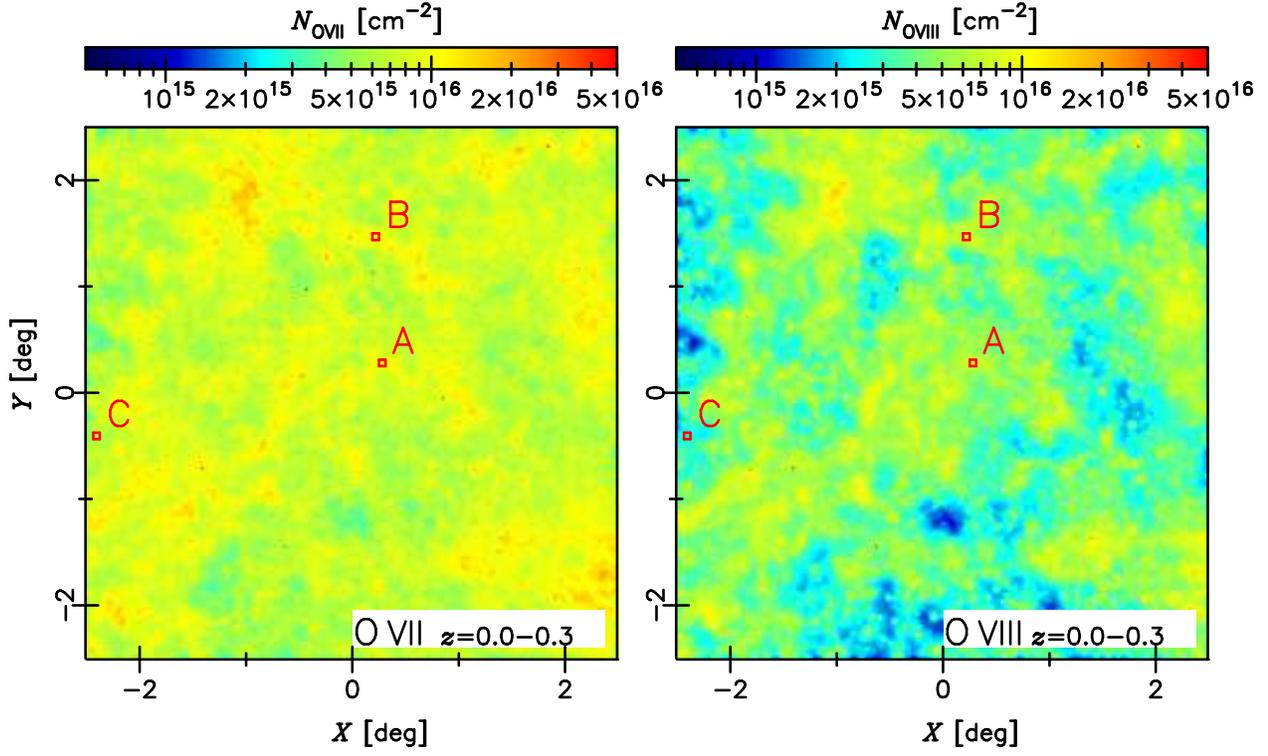

 \begin{center}
 \rotatebox[origin=c]{270}{
  \FigureFile(100mm,100mm){figure7.eps}}
  \end{center}
 %\vspace{-0.5cm}
 \caption{Column density maps of O{\sc vii} ({\it Left}) and
 O{\sc viii} ({\it Right}) for $0.0 < z < 0.3$. Three red squares
 labeled A,  B, and C indicate the regions corresponding to Figures 
 \ref{fig:mock_spectra} to \ref{fig:t_delta}.
 \label{fig:columndens_map}}
\end{figure}

\begin{figure}
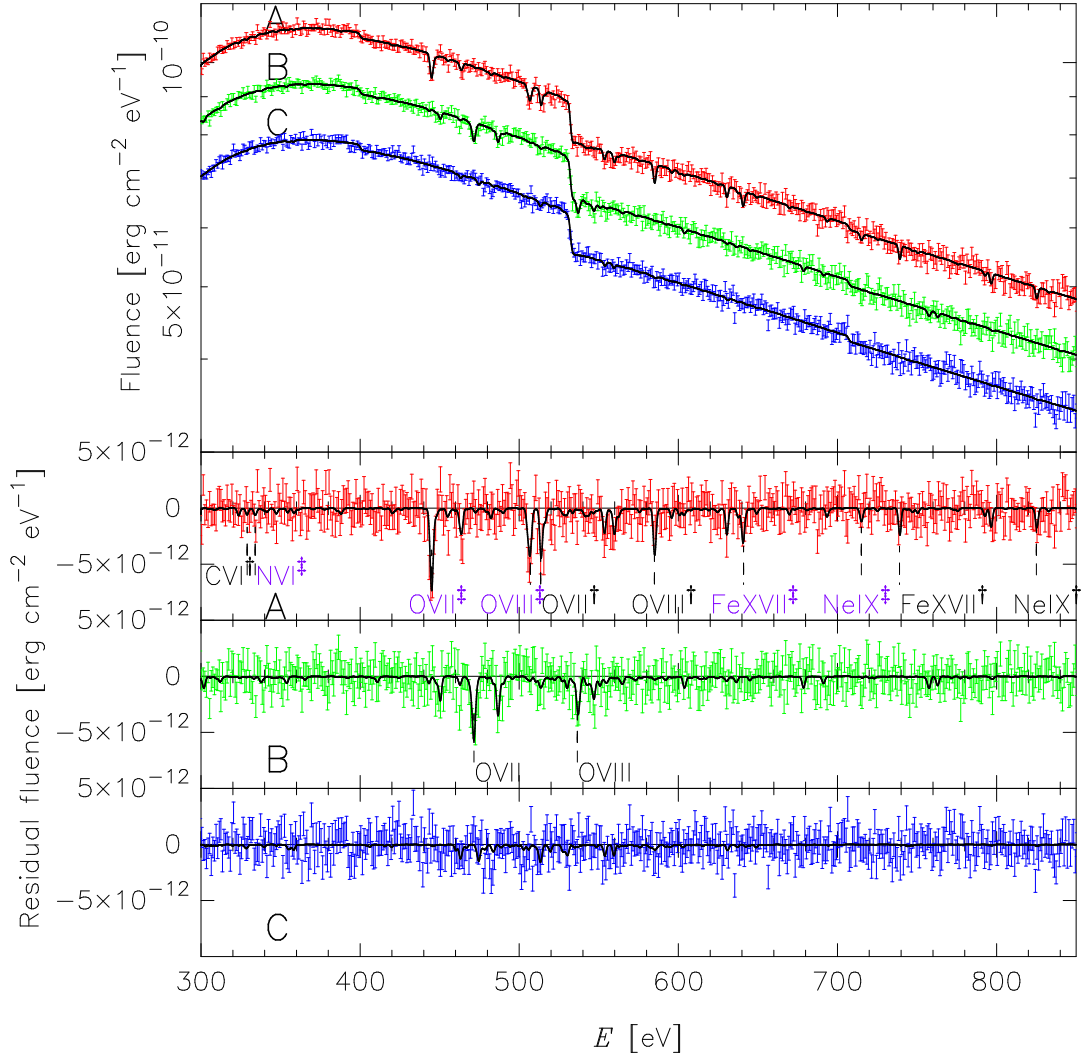

\begin{center}
  \rotatebox[origin=c]{270}{
  \FigureFile(140mm,300mm){figure8.eps}
 }
\end{center}
 \caption{Mock transmission spectra of a QSO.
Top panel plots
fluences through three different LOSs (A, B, and C indicated in 
 Figure \ref{fig:columndens_map}). For an illustrative purpose, curves B
 and C are artificially multiplied by a factor of $10^{-0.2}$ and
 $10^{-0.4}$.  Lower panels show their residual fluence normalized by the
 continuum level.  The metal line systems with labels $\dagger$ and
 $\ddagger$ in the spectrum A correspond to the WHIM clumps at $z=0.12$
 ({\it Upper panels} in Fig.\ref{fig:map12}) and $z=0.29$ ({\it Lower
 panels} in Fig.\ref{fig:map12}),
 respectively. \label{fig:mock_spectra}}
\end{figure}

\begin{figure}[htbp]
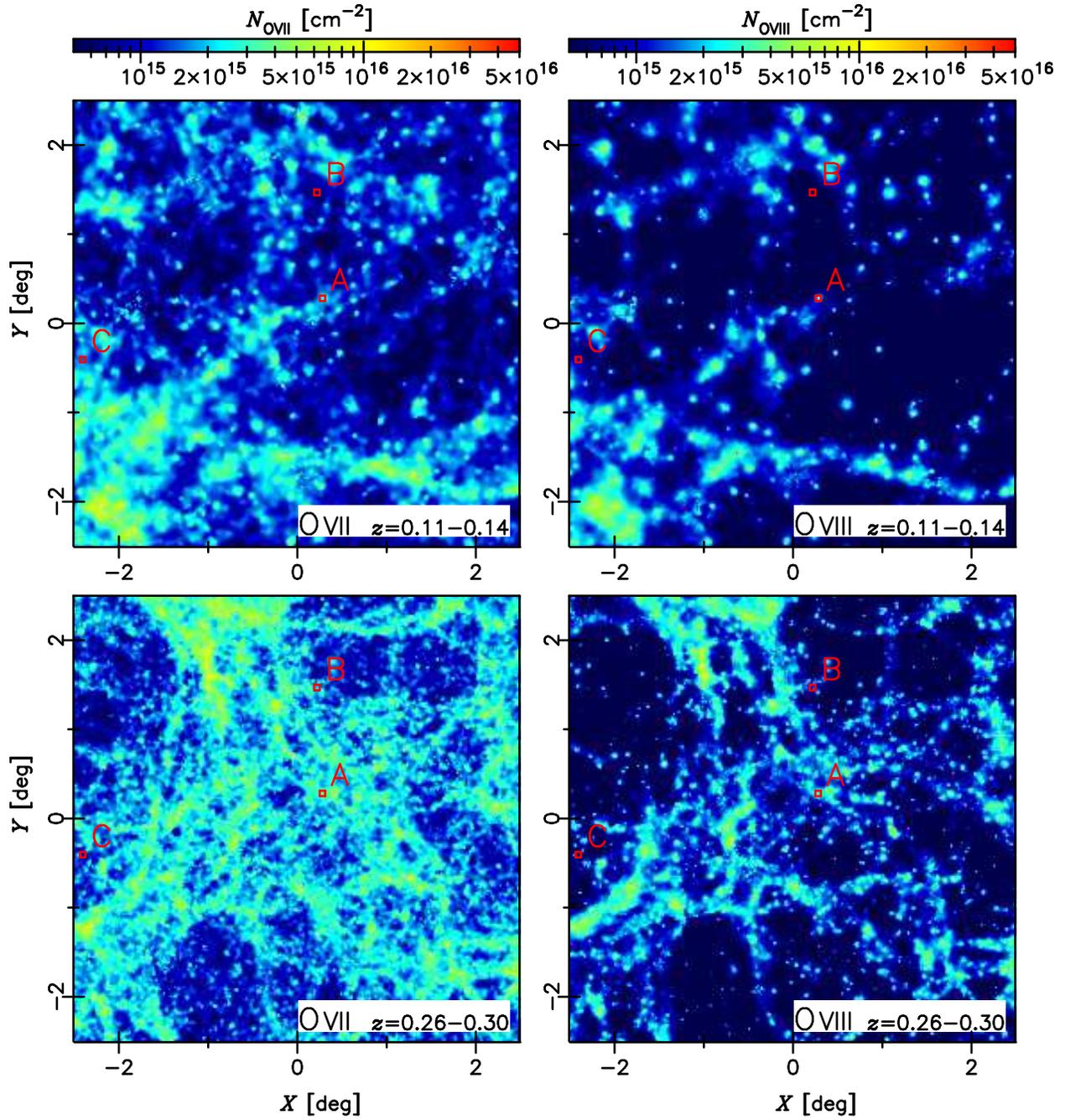

 \begin{center}
 \rotatebox[origin=c]{270}{
  \FigureFile(180mm,180mm){figure9.eps}}
 \end{center}
 \vspace{-0.5cm}
 \caption{Column density maps of O{\sc vii} ({\it Left}) and O{\sc viii}
 ({\it Right}) for for $0.11 < z < 0.14$ ({\it Upper}) and for
 $0.26 < z < 0.30$ ({\it Lower}).  \label{fig:map12}}
\end{figure}

\begin{figure}[htbp]
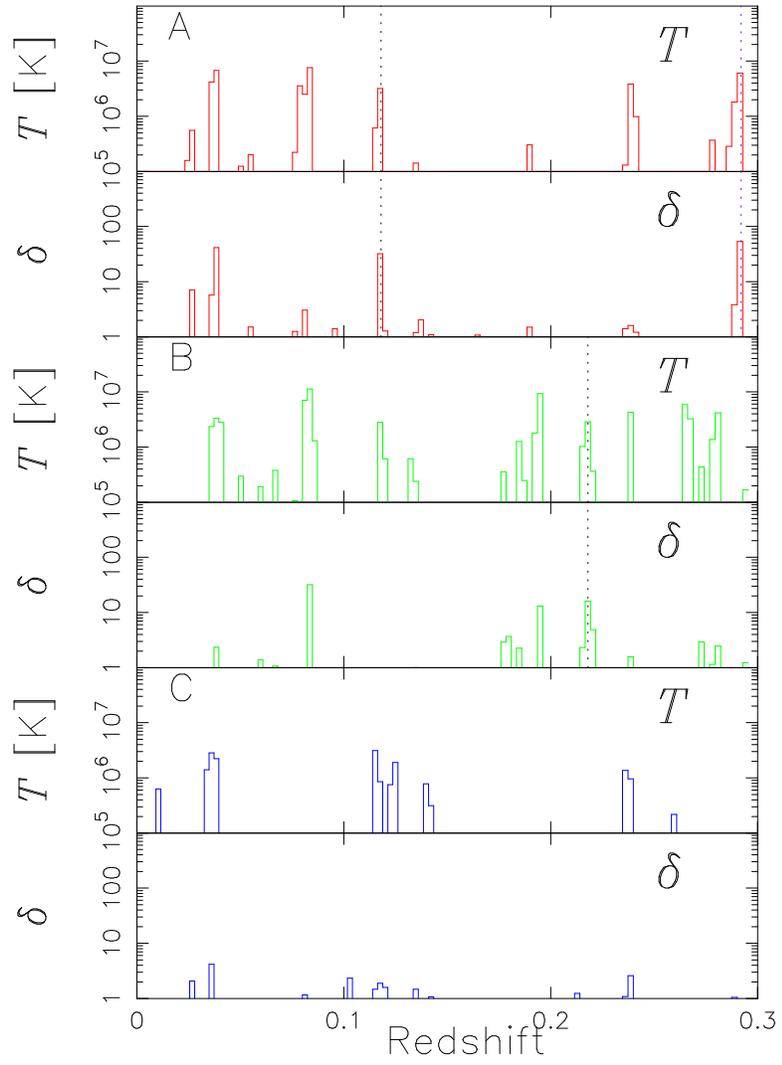

 \begin{center}
 \rotatebox[origin=c]{270}{
  \FigureFile(140mm,140mm){figure10.eps}}
 \end{center}
 \caption{Redshift distribution of temperature and over-density along
 the three LOSs A, B, and C (from top to bottom). \label{fig:t_delta}}
\end{figure}

\begin{figure}[htbp]
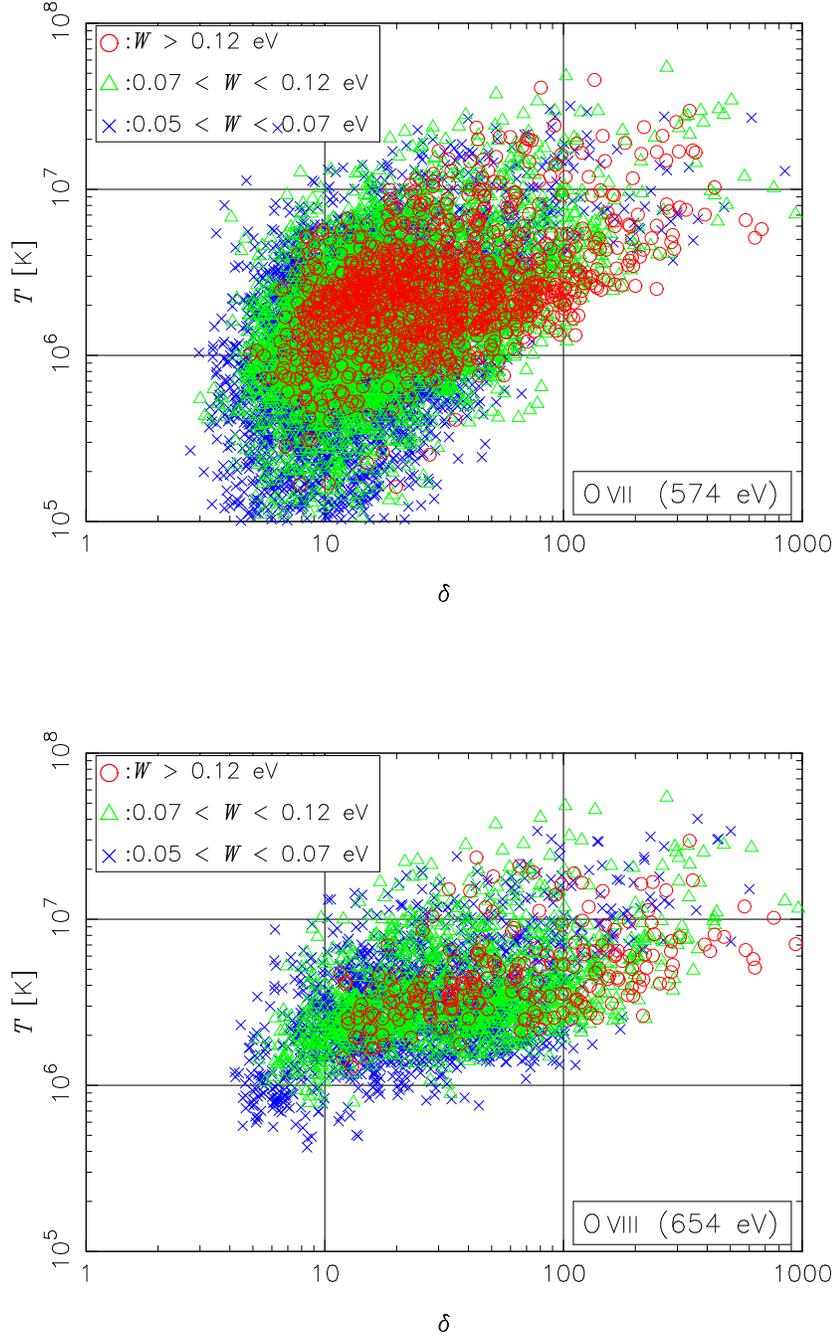

 \begin{center}
 \rotatebox[origin=c]{270}{
  \FigureFile(80mm,80mm){figure11a.eps}}
 \rotatebox[origin=c]{270}{
  \FigureFile(80mm,80mm){figure11b.eps}}
 \end{center}
 \caption{Scatter plots of over-density and temperature of
 absorption line systems; O{\sc vii} 574 eV ({\it
 Upper panel}) and O {\sc viii} 654 eV ({\it Lower panel}).
 Circles, crosses, and triangles correspond to
the systems with equivalent width of $W \ge 0.12$ eV, $0.07 \le W \le
 0.12$ eV, and $0.05 \le W \le 0.07$ eV, respectively. 
\label{fig:rho_T}}
\end{figure}

\begin{figure}[htbp]
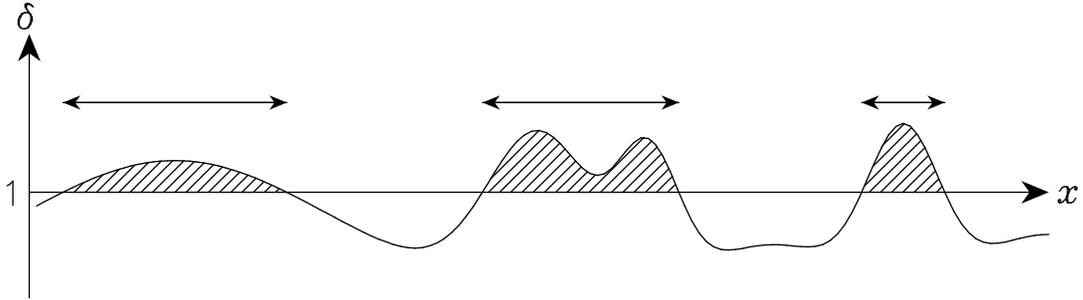

 \begin{center}
 \rotatebox[origin=c]{270}{
 \FigureFile(40mm,40mm){figure12.eps}}
 \caption{Schematic identification method of clumps of WHIM. 
 \label{fig:identify_whim}}
 \end{center}
\end{figure}

\begin{figure}[htbp]
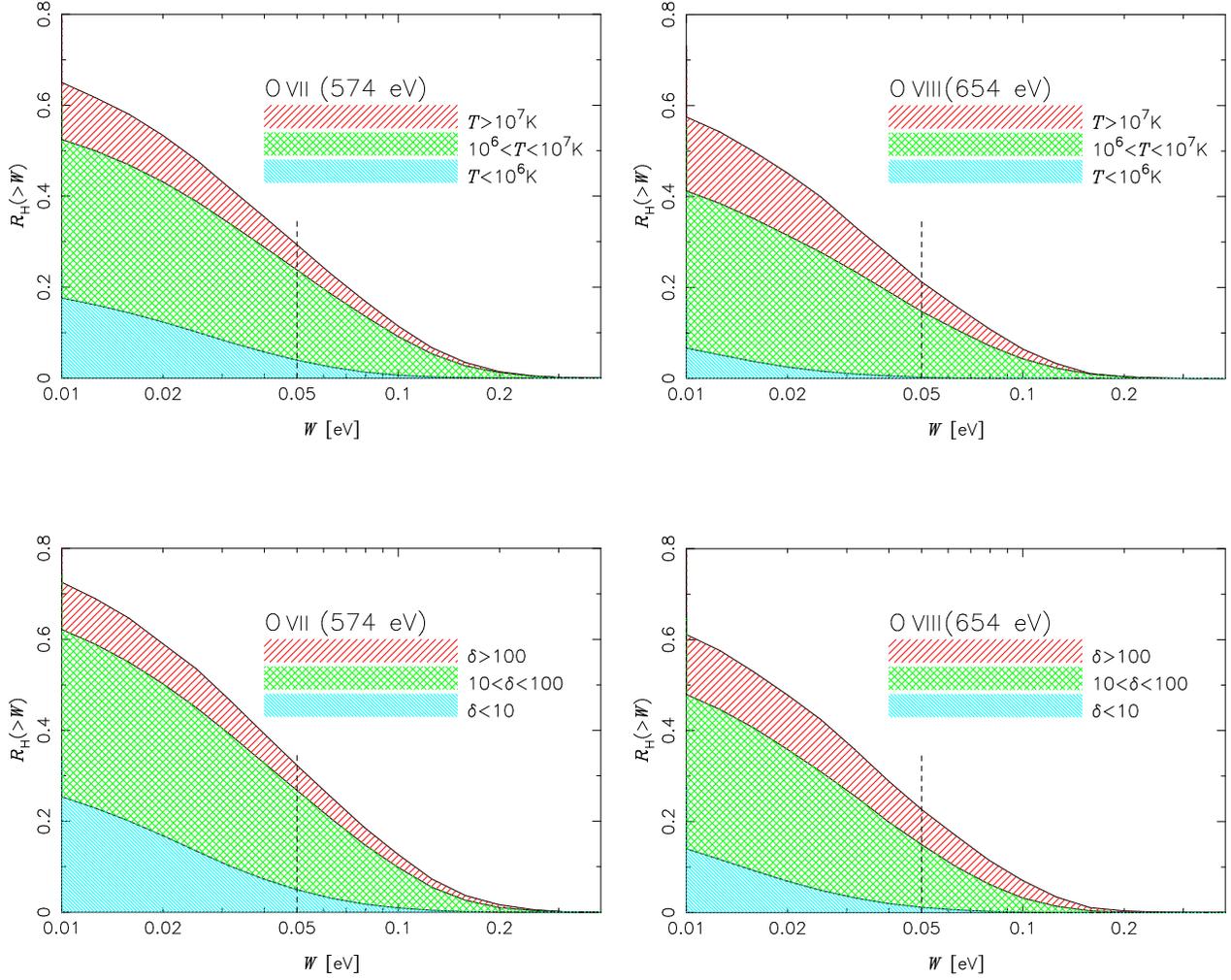

 \begin{center}
\rotatebox[origin=c]{270}{
 \FigureFile(60mm,60mm){figure13a.eps}}
 \rotatebox[origin=c]{270}{
  \FigureFile(60mm,60mm){figure13b.eps}}
\rotatebox[origin=c]{270}{
 \FigureFile(60mm,60mm){figure13c.eps}}
 \rotatebox[origin=c]{270}{
  \FigureFile(60mm,60mm){figure13d.eps}}
 \end{center}
 \caption{Cumulative ratio of hydrogen column densities
 (eq.(\ref{eq:r_h_w})) as a function of the equivalent width of the
 detected absorption line systems; O{\sc vii} 574 eV ({\it Left})
 and O {\sc viii} 654 eV ({\it Right}).  The three shaded regions
 indicate the different temperature ({\it Upper}) and density ({\it
 Lower}) ranges; $T>10^7$ K, $10^6<T<10^7$ K, and $T<10^5$K ({\it
 Upper}), and $\delta>100, 10<\delta<100$, and $\delta<10$ ({\it Lower}) 
 from top to bottom.  \label{fig:r_h_w}}
\end{figure}

\begin{figure}[htbp]
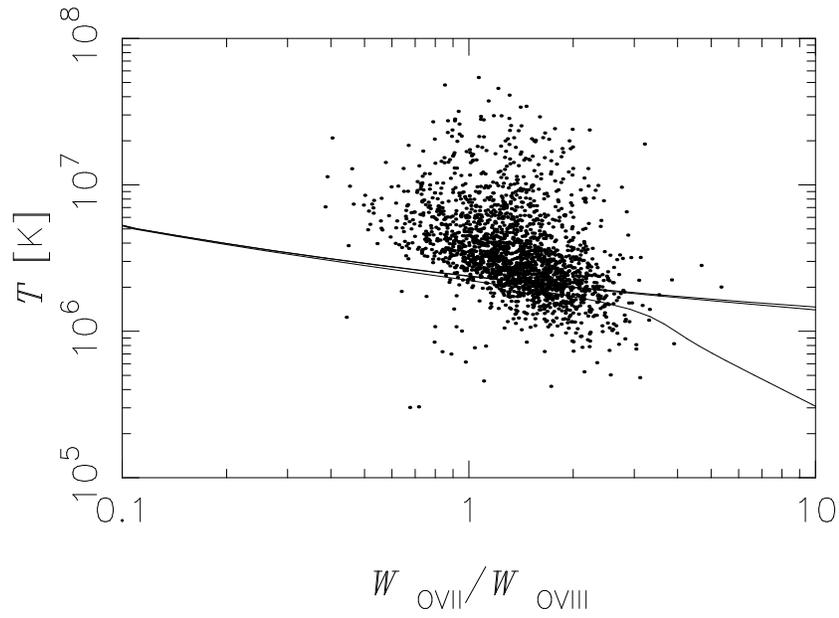

 \begin{center}
 \rotatebox[origin=c]{270}{
  \FigureFile(80mm,80mm){figure14.eps}}
 \end{center}
 \caption{Ratio of the O{\sc vii} (574 eV) and O {\sc viii} (654 eV)
 equivalent widths plotted against the mass-weighted average
 temperature. The solid curves show the theoretical curves in the
 uniform case under CPIE assuming the hydrogen number density $n_H =
 10^{-5}, 10^{-6}$, and $10^{-7} \mathrm{cm^{-3}}$ from top to
 bottom.\label{fig:scat_temp}}
\end{figure}

\begin{figure}[htbp]
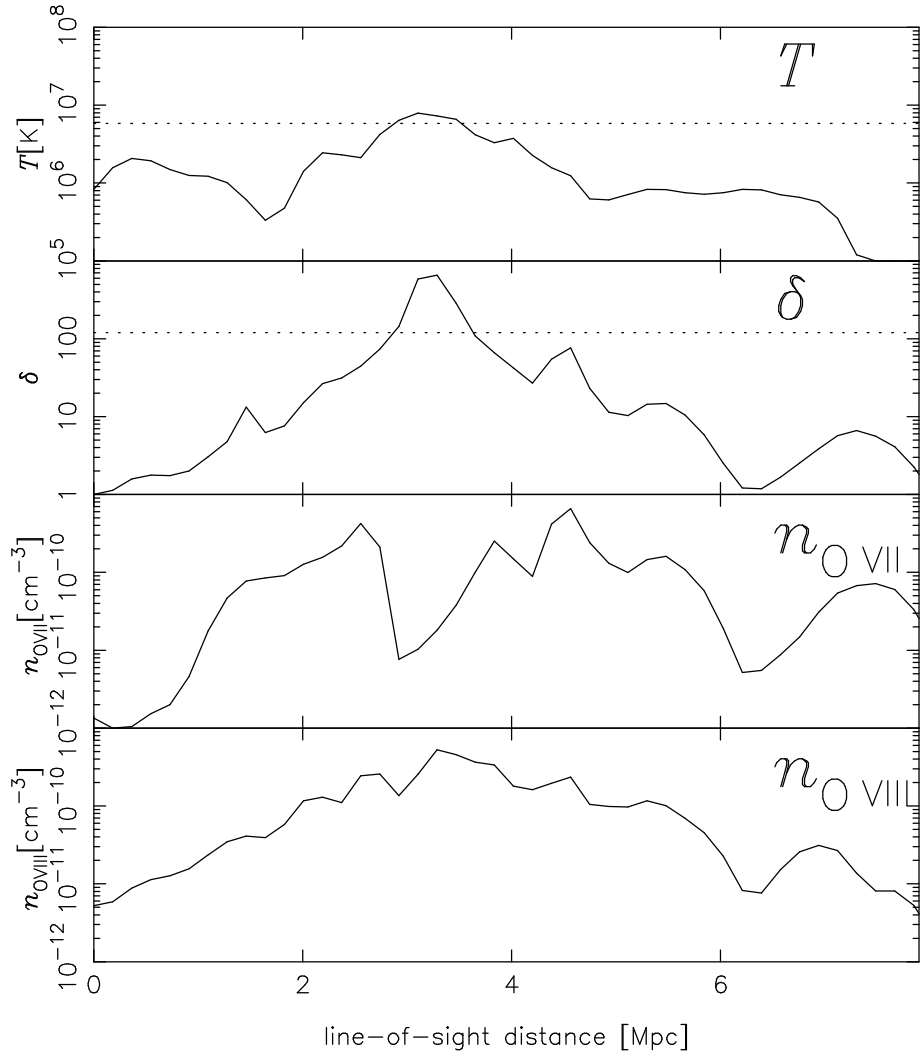

 \begin{center}
 \rotatebox[origin=c]{270}{
  \FigureFile(140mm,140mm){figure15.eps}}
 \end{center}
\caption{An example of sub-structure in a WHIM which forms a single
 absorption line system according to our identification procedure. From
 top to bottom plotted are temperature, over-density, number density of
 O {\sc vii}, and number density of O {\sc viii}.
\label{fig:prof}}
\end{figure}

\begin{figure}[htbp]
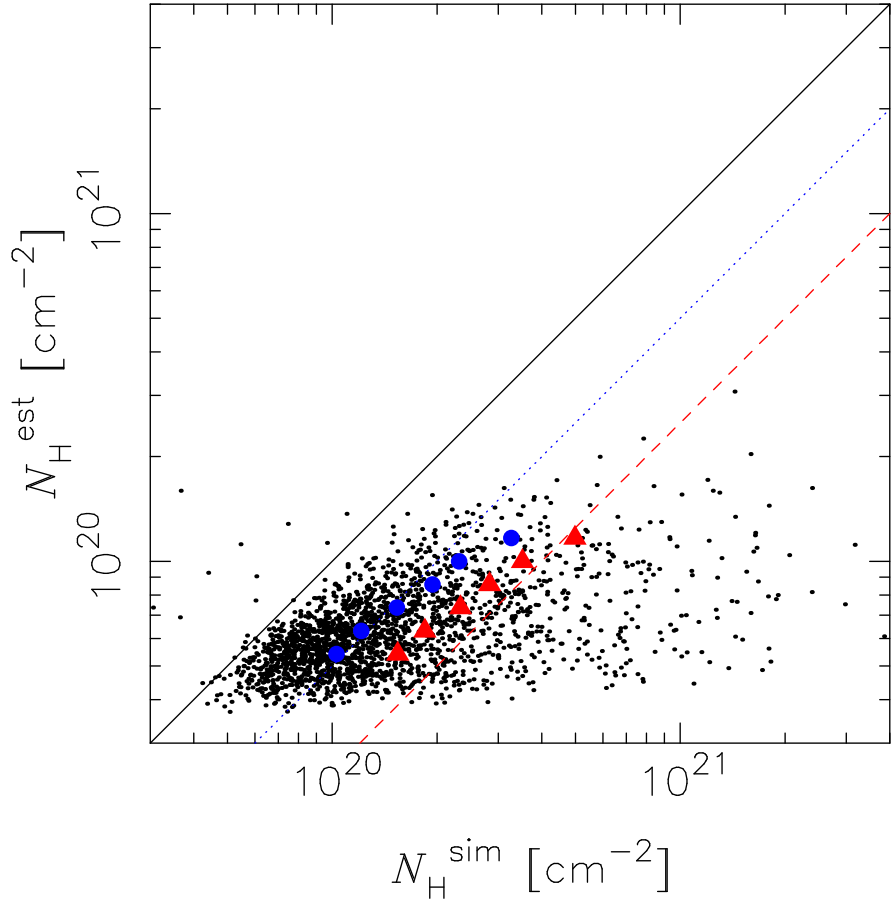

 \begin{center}
 \rotatebox[origin=c]{270}{
  \FigureFile(120mm,120mm){figure16.eps}}
 \end{center}
 \caption{Comparison between $N_{H}^{\mathrm{est}}(>3\sigma)$ and
$N_{H}^{\mathrm{sim}}(>3\sigma)$. Each dot represents a single WHIM
clump which exhibits both O{\sc vii} and O{\sc viii} with $S/N>3$ ($1903$
clumps in total).  The filled circles and triangles indicate the mean
and median values of $N_{H}^{\mathrm{sim}}(>3\sigma)$ for the fixed bin
with respect to $N_{H}^{\mathrm{est}}(>3\sigma)$. The solid, dotted, and
dashed lines show
$N_{H}^{\mathrm{est}}(>3\sigma)/N_{H}^{\mathrm{sim}}(>3\sigma)=1$,
$1/2$, and $1/4$, respectively.  \label{fig:nhest-nhsim}}
\end{figure}

\begin{figure}[htbp]
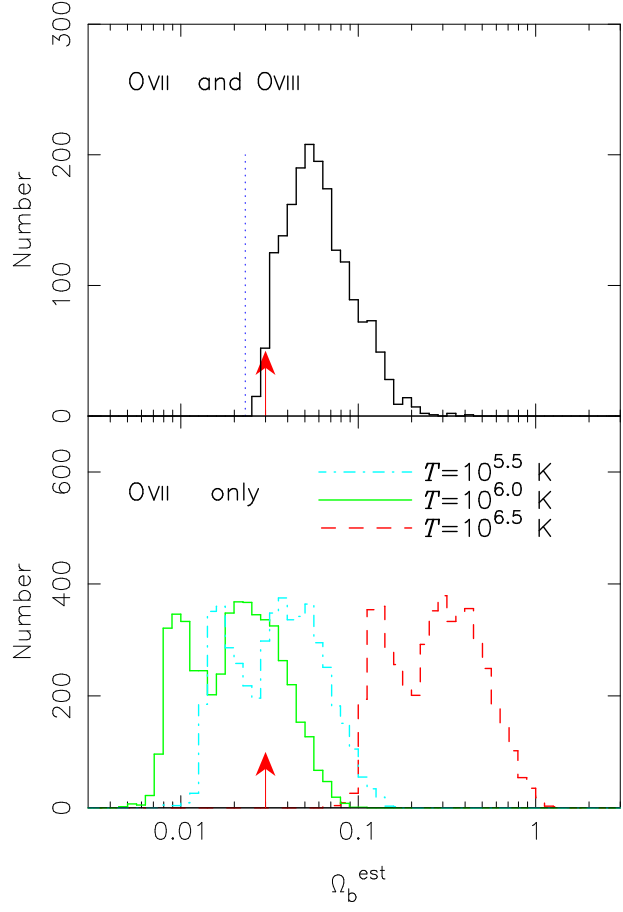

 \begin{center}
 \rotatebox[origin=c]{270}{
  \FigureFile(120mm,120mm){figure17.eps}}
 \end{center}
 \caption{Histogram of $\Omega_b^{\mathrm{est}}$ estimated according to
eq.(\ref{eq:omega_est}). The upper panel shows the case that both O{\sc
vii} and O{\sc viii} are detected and temperature is estimated from
their line ratio. The vertical dotted line is the mean value $0.023$
computed from the entire LOSs including non-detection.  The lower panel
shows an estimation using only O {\sc vii} absorption lines assuming
temperature $T=10^{5.5}$, $10^{6.0}$, and $10^{6.5}$ K. The arrow
indicates the value of $\Omega_b (=0.03)$ that is adopted in our
numerical simulation.  \label{fig:omega}}
\end{figure}

In passing we have to emphasize once again that our simulation data
adopted $\Omega_{\rm b}=0.03$ rather than the currently more favored
value $0.04$. Also we restrict our current analysis up to $z=0.3$ where
a spectrum is relatively free from the Galactic confusion, but in
principle one can explore WHIM systems at higher redshifts ($z\sim 1$)
with QSO(GRB).  Therefore while our present conclusions
concerning the 
detectability of WHIM in absorption toward a QSO(GRB) may seem fairly
modest,  
most likely they are very conservative. Moreover the result is very
sensitive to the metallicity of WHIM which is rather uncertain and is
likely to significantly vary from place to place. Given the strong
metallicity dependence as shown in Figure \ref{fig:metal_dep}, what is
relevant is not the mean metallicity of WHIM but the fraction of WHIM
clumps which exhibit detectable equivalent width. In this respect, the
wide variation of the metallicity, even if its overall mean is
$0.1Z_\odot$ as we most adopted here, is expected to {\it
systematically} increase the detectability.

Throughout this paper, we adopted $\Delta E = 2 {\rm eV}$
with calorimeter on-board {\it XEUS} in mind.
Current observations of WHIM by several grating detectors are 
indicative. With XEUS's large effective area, WHIM absorption features 
will be collected with a much larger number of samples as a 
by-product of QSO studies. With careful statical studies as we presented 
in this paper, these significantly contribute to the proper understanding 
of physics of WHIM.
If higher energy resolution with 
gratings, for example, $\Delta E=0.15 {\rm eV}$ at $E=0.6 {\rm keV}$,
becomes available in future, it is even possible to obtain line profile
accurately and estimate the effect of turbulence.  Also one can study
multiphase nature in a single absorption system.
Furthermore, since signal-to-noise ratio
improves as $1/\sqrt{\Delta E}$, one can significantly increase the
number of detectable oxygen and other ion absorption systems.  These
significantly contribute to the proper understanding of
physics of WHIM \citep{pharos}.

In order to be more quantitative, we need much more advanced simulation
datasets which include more realistic physical processes such as
radiative cooling, photo-ionization, star formation, supernova feedback,
and metallicity evolution. This is exactly what we are performing, and
we hope to revisit the above issues in due course.

\medskip 

Numerical computations presented in this paper were carried out at ADAC
(the Astronomical Data Analysis Center) of the National Astronomical
Observatory, Japan (project ID: mky05a). 
We thank the referee Fabrizio Nicastro for very constructive
and useful comments. We believe that
his numerous suggestions significantly improve the present paper.
We are also grateful to Kei Hiraki and Mari Inaba for their generous
allocation of computer 
resources at the University of Tokyo supported by the Special
Coordination Fund for Promoting Science and Technology, Ministry of
Education, Culture, Sport, Science and Technology.  This work was
supported in part by Grants-in-Aid for Scientific Research from the
Japan Society for Promotion of Science (Nos.14102004, 14204017,
15340088, 15740157, and 16340053).

\newpage
\onecolumn

%-----------------------------------------------------------------------
\begin{table}[h]
  \caption{Ten prominent resonant absorption lines in 0.3 -- 2.0 keV}  
\label{tab:abs_species}
 \begin{center}
  \begin{tabular}{lccccc}
   \hline\hline
   \multicolumn{1}{c}{species} & energy& oscillator strength \\
   \hline
   C{\sc vi} & 368 eV & 0.416 \\
   \hline
   N{\sc vi} & 431 eV & 0.675 \\
   N{\sc vii} & 500 eV & 0.416 \\
   \hline
   O{\sc vii} & 574 eV & 0.696 \\
   O{\sc vii} & 666 eV & 0.146 \\
   O{\sc viii}  & 654 eV & 0.416  \\
   \hline
   Ne{\sc ix} & 922 eV & 0.724 \\
   Ne{\sc x} & 1022 eV & 0.416 \\
   \hline
   Mg{\sc xi}& 1352 eV& 0.742 \\
   \hline
   Fe{\sc xvii} & 826 eV & 2.96 \\
   \hline
  \end{tabular}
 \end{center}
\end{table}
%-----------------------------------------------------------------------

%%-----------------------------------------------------------------------
\begin{table}[h]
  \caption{Expected number of oxygen absorption line systems with
 $S/N \ge  3$ and  $S/N \ge 2$ per LOS up to $z=0.3$
 toward a QSO discussed in \S 2.3. \label{tab:abs_oxy}} 
 \begin{center}
  \begin{tabular}{ccccc}
   \hline\hline
   \multicolumn{2}{c}{} & \multicolumn{3}{c}{expected number/LOS} \\
      \multicolumn{1}{c}{metallicity model} & \multicolumn{1}{c}{}& \multicolumn{1}{c}{O{\sc
   vii} (574 eV)}&
   \multicolumn{1}{c}{O{\sc viii} (654 eV)} & \multicolumn{1}{c}{O{\sc
   vii} and O{\sc viii}}\\
   \hline
$Z = 0.1 Z_{\odot}$  & $S/N \ge  3$ & 1.58 & 0.37 & 0.35\\
 & $S/N \ge  2$ &  3.07 & 0.82 & 0.79\\
\hline
$ Z = {\rm min}[0.2,0.02 (\rho/\bar{\rho})^{0.3}] Z_{\odot}$ & $S/N \ge
   3$ & 0.50 & 0.15 & 0.14  \\
 & $S/N \ge  2$ & 0.97 & 0.31 & 0.28\\
\hline
$Z = 0.3 Z_{\odot}$  & $S/N \ge 3$ & 6.71 & 2.19 & 2.10 \\
 & $S/N \ge 2$ &9.91  & 3.40 & 3.22\\
\hline
  \end{tabular}
 \end{center}
\end{table}
%%-----------------------------------------------------------------------

%%-----------------------------------------------------------------------
\begin{table}[h]
  \caption{Expected number of oxygen absorption line systems with
 $S/N \ge  3$ and  $S/N \ge 2$ per LOS up to
 $z=0.3$ toward a GRB afterglow discussed in \S
 2.4. \label{tab:abs_oxy_grb}}    
 \begin{center}
  \begin{tabular}{ccccc}
   \hline\hline
   \multicolumn{2}{c}{} & \multicolumn{3}{c}{expected number/LOS} \\
      \multicolumn{1}{c}{metallicity model} & \multicolumn{1}{c}{}& \multicolumn{1}{c}{O{\sc
   vii} (574 eV)}&
   \multicolumn{1}{c}{O{\sc viii} (654 eV)} & \multicolumn{1}{c}{O{\sc
   vii} and O{\sc viii}}\\
   \hline
$Z = 0.1 Z_{\odot}$  & $S/N \ge  3$ & 1.71 & 0.43 & 0.41\\
 & $S/N \ge  2$ &  3.30 & 0.94 & 0.90\\
\hline
$ Z = {\rm min}[0.2,0.02 (\rho/\bar{\rho})^{0.3}] Z_{\odot}$ & $S/N \ge
   3$ & 0.55 & 0.17 & 0.15  \\
 & $S/N \ge  2$ & 1.05 & 0.35 & 0.32\\
\hline
$Z = 0.3 Z_{\odot}$  & $S/N \ge 3$ & 7.08 & 2.39 & 2.28 \\
 & $S/N \ge 2$ &10.3  & 3.66 & 3.45\\
\hline
  \end{tabular}
 \end{center}
\end{table}
%%-----------------------------------------------------------------------

\clearpage
%%%%%%%%%%%%%%%%%%%%%%%%%%%%%%%%%%%%%%%%%%%%%%%%%%%%%%%%%%%%%%%%%%%%%%%%

%%-----------------------------------------------------------------------
%%-----------------------------------------------------------------------

%%-----------------------------------------------------------------------
%%-----------------------------------------------------------------------

%%-----------------------------------------------------------------------
%%-----------------------------------------------------------------------

%%-----------------------------------------------------------------------
%%-----------------------------------------------------------------------

%%-----------------------------------------------------------------------
%%-----------------------------------------------------------------------

%%-----------------------------------------------------------------------
%%-----------------------------------------------------------------------

%%-----------------------------------------------------------------------
%%-----------------------------------------------------------------------

%%-----------------------------------------------------------------------
%%-----------------------------------------------------------------------

%%-----------------------------------------------------------------------
%%-----------------------------------------------------------------------

%%-----------------------------------------------------------------------
%%-----------------------------------------------------------------------

%%-----------------------------------------------------------------------
%%-----------------------------------------------------------------------

%%-----------------------------------------------------------------------
%%-----------------------------------------------------------------------

%%-----------------------------------------------------------------------
%%-----------------------------------------------------------------------

%%-----------------------------------------------------------------------
%%-----------------------------------------------------------------------

%%-----------------------------------------------------------------------
%%-----------------------------------------------------------------------

%%-----------------------------------------------------------------------
%%-----------------------------------------------------------------------

%%-----------------------------------------------------------------------
%%-----------------------------------------------------------------------


\bigskip 

\begin{thebibliography}{}
 \bibitem [Aguirre et al.(2001)]{Aguirre2001} Aguirre, A., Hernquist,
          L., Schaye, J., Katz, N., Weinberg, D.H., \& Gardner, J. 2001,
          ApJ, 561, 521
 \bibitem [Anders \& Grevesse(1989)]{Anders1989} Anders, E., \&
	  Grevesse, N. 1989, Geochim. Cosmochim. Acta, 53, 197
 \bibitem [Brinkmann, Yuan, Siebert(1997)]{Brinkmann1997} Brinkmann, W.,
	  Yuan, W., \& Siebert, J. 1997, A\&A, 319, 413
 \bibitem [Cen \& Fang(2006)]{CenFang} Cen, R. \& Fang, T. 2006,
          ApJ, submitted (astro-ph/0601009)
 \bibitem [Cen \& Ostriker(1999a)]{Cen1999a} Cen, R. \& Ostriker, J. 1999a,
          ApJ, 514, 1
 \bibitem [Cen \& Ostriker(1999b)]{Cen1999b} Cen, R. \& Ostriker, J. 1999b,
          ApJ, 519, L109
 \bibitem [Chen et al.(2003)]{Chen2003} Chen, X.,Weinberg, D.H., \& Dav\'e,
	  R. 2003, ApJ, 594, 42 
 \bibitem [Dav\'e et al.(2001)]{Dave2001} Dav\'e, R., Cen, R., Ostriker,
          J.P., Bryan, G.L., Hernquist, L., Katz, N., Weinberg, D.H.,
          Norman, M.L., \& O'Shea, B. 2001, ApJ, 552, 473
 \bibitem [De Pasquale et al.(2005)]{DePasquale} De Pasquale, M., Piro,
	  L., Gendre, B., Amati, L., Antonelli, L.A., Costa, E., Feroci,
	  M., Frontera, F., Nicastro, L., Soffitta, P., \& in't Zand,
	  J. 2005, astro-ph/0507708
 \bibitem [Fang, Bryan, Canizares(2002)]{Fang2002a} Fang, T., Bryan,
          G.L., \& Canizares, C.R. 2002, ApJ, 564, 604
 \bibitem [Fang et al.(2002)]{Fang2002b} Fang, T., Marshall, H.L., Lee,
          J.C., Davis, D.S., \& Canizares, C.R., 2002, ApJ, 572, L127
 \bibitem [Fiore et al.(2000)]{Fiore2000} Fiore, F., Nicastro, F.,
	  Savaglio, S., Stella, L., \& Vietri, M., 2000, ApJ, 544, L7
 \bibitem [Fukugita, Hogan, \& Peebles(1998)]{Fukugita1998} Fukugita,
          M., Hogan, C.J., \& Peebles, P.J.E. 1998, ApJ, 503, 518
 \bibitem [Fukugita \& Peebles(2004)]{Fukugita2004} Fukugita,
          M., \& Peebles, P.J.E. 2004, ApJ, 616, 643
 \bibitem [Fujimoto et al.(2004)]{Fujimoto2004} Fujimoto, R., Takei, Y.,
	  Tamura, T., Mitsuda, K., Yamasaki, N.Y., Shibata, R., Ohashi,
	  T., Ota, N., Audley, 
	  M.D., Kelley, R.L., Kilbourne, C.A., 2004,
	  PASJ, 56, L29
 \bibitem [Jager et al 1997]{Jager1997} 
	  Jager, R., Mels, W. A., Brinkman, A. C., Galama, M. Y.,
	  Goulooze, H., Heise, J., Lowes, P., Muller, J. M., Naber, A.,
	  Rook, A., Schuurhof, R., Schuurmans, J. J., Wiersma, G., 1997,
	  A\&A S, 125, 557
 \bibitem [Kang et al.(2005)]{Kang2005} Kang, H., Ryu, D., Cen, R.,  Song,
          D., 2005, ApJ, 620, 21
 \bibitem [Morrison \& McCammon(1983)]{Morrison1983} 
          Morrison R. \& McCammon D., 1983, ApJ, 270, 119 
 \bibitem [Miyaji et al.(1998)]{Miyaji1998} Miyaji, T., Ishisaki, Y.,
          Ogasaka, Y., Ueda, Y., Freyberg, M. J., Hasinger, G., \&
          Tanaka, Y. 1998, A\&A, 334, 13
 \bibitem [Mathur et al.(2003)]{Mathur2003} Mathur, S., Weinberg, D.H.,
	  \& Chen, X. 2003, ApJ, 582, 82  
 \bibitem [Nicastro(2005)]{pharos} Nicastro, F. 2005, talk presented at
	  an international workshop on {\it Measuring the Diffuse
	  Intergalactic Medium}, Kanagawa, Japan.
 \bibitem [Nicastro et al.(2002)]{Nicastro2002} Nicastro, F., Zezas, A.,
          Drake, J., Elvis, M., Fiore, F., Fruscione, A., Marengo, M.,
          Mathur, S., \& Bianchi, S. 2002, ApJ, 573, 157
 \bibitem [Nicastro et al.(2005a)]{Nicastro2005} Nicastro, F., Mathur,
	  S., Elvis, M., Drake, J., Fang, T., Fruscione, A., Krongold,
	  Y., Marshall, H., Williams, \&  R., Zezas, A. 2005a, Nature,
	  433, 495
 \bibitem[Nicastro et al.(2005b)]{Nicastro2005b} Nicastro, F., Mathur,
	  S., Elvis, M., Drake, J., Fiore, F., Fang, T., Frauscione, A.,
	  Krongold, Y., Marshall, H., \& Williams, R. 2005b, ApJ, 629,
	  718 
 \bibitem [Ohashi et al.(2004)]{Ohashi2004} Ohashi, T. et al., 2003, 
	  in Proceedings of ``Modelling the Intergalactic and
	  Intracluster Media'' (astro-ph/0402546)
 \bibitem [Piro (2004)]{Piro2004} Piro, L., 2004, 
	  in Proceedings of ``Third Rome Workshop on Gamma-Ray Bursts in
	  the Afterglow Era'' (astro-ph/0402638)   
 \bibitem [Shull et al.(1999)]{Shull1999} Shull, J.M., Roberts, D.,
	  Giroux, M.L., Penton, S.V., \& Fardal, M.A. 1999, AJ, 118, 1450 
 \bibitem [Spergel et al.(2003)]{Spergel03} Spergel, D. N. et al.  2003,
	  ApJS, 148, 175
%%% 2006/06/10 deleted
% \bibitem [Stratta et al.(2004)]{Stratta2004} Stratta, G., Fiore, F.,
%	  Antonelli, L.A., Piro, L. \& De Pasquale, M. 2004,
%	  ApJ, 608, 846 
 \bibitem [Suto et al.(2004a)]{Suto2004a} Suto, Y., Yoshikawa, K.,
	  Yamasaki, N.Y., Mitsuda, K., Fujimoto, R., Furusho, T.,
	  Ohashi, T., Ishida, M., Sasaki, S., Ishisaki, Y., Tawara, Y.,
	  Furuzawa, A. 2004a, Journal of the Korean Physical Society,
	  45, S110 
 \bibitem [Suto et al.(2004b)]{Suto2004b} Suto, Y., Yoshikawa, K., Dolag,
	  K., Sasaki, S., Yamasaki, N.Y., Ohashi, T., Mitsuda, K.,
	  Tawara, Y., Fujimoto, R., Furusho, T., Furuzawa, A., Ishida,
	  M., Ishisaki, Y., Takei, Y.  2004b, Journal of the Korean
	  Astronomical Society, 37, 387 
 \bibitem [Tripp, Savage \& Jenkins(2000)]{Tripp2000} Tripp, T.M.,
          Savage, B.D., \& Jenkins, E.B. 2000, ApJ, 534, L1
 \bibitem [Verner et al.(1996)]{Verner1996} Verner, D.A., Verner, E.M., \&
	  Ferland, G.J. 1996, Atomic Data Nucl. Data Table, 64, 1
 \bibitem [Yoshikawa et al.(2001)]{Yoshikawa2001} Yoshikawa, K., Taruya, A.,
         Jing, Y.P., \& Suto, Y. \ 2001, ApJ, 558, 520
 \bibitem [Yoshikawa et al.(2003)]{Yoshikawa2003} Yoshikawa, K.,
	  Yamasaki, N.Y., Suto, Y., Ohashi, T., Mitsuda, K., Tawara,
	  Y., Furuzawa, A. \ 2003, PASJ, 55, 879 (Paper I) 
 \bibitem [Yoshikawa et al.(2004)]{Yoshikawa2004} Yoshikawa, K.,
	  Dolag, K., Suto, Y., Sasaki, S., Yamasaki, N.Y., Ohashi, T.,
	  Mitsuda, K., Tawara, Y., Fujimoto, R., Furusho, T., Furuzawa,
	  A., Ishida, M., Ishisaki, Y., \& Takei, Y. \ 2004, PASJ, 56,
	  939 (Paper II)
 \bibitem [Yoshikawa, Sasaki (2006)]{YoshikawaSasaki} 
   Yoshikawa, K., \& Sasaki, S. \ 2006, PASJ in press. (astro-ph/060372)
 \bibitem [Yuan et al.(1998)]{Yuan1998} Yuan, W., Brinkmann, W.,
	  Siebert, J., \& Voges, W. \ 1998, A\&A, 330, 108 
\end{thebibliography}
\end{document}